\documentclass[10pt]{article}
\usepackage{array,multirow}
\usepackage[utf8]{inputenc}
\usepackage[linesnumbered,ruled]{algorithm2e}
\usepackage{amsmath}
\usepackage{graphicx}
\usepackage{caption}
\usepackage{subcaption}
\usepackage{datetime}
\usepackage{longtable}
\usepackage{booktabs}
\usepackage{lscape}
\usepackage{hyperref}
\hypersetup{
    colorlinks = true,
    allcolors = blue
}
\usepackage{times}
\usepackage{float}
\restylefloat{table}

\usepackage[margin=1in]{geometry}

\usepackage[parfill]{parskip}
\usepackage{setspace}
\usepackage[title]{appendix}

\usepackage[dvipsnames]{xcolor}

\usepackage{enotez} 

\usepackage{amssymb}

\usepackage[inline]{enumitem}

\usepackage[export]{adjustbox}

\usepackage[round,authoryear]{natbib}
 \setcitestyle{authoryear,open={(},close={)}}
\usepackage{authblk}
\usepackage{tikz}
\usetikzlibrary{decorations.pathreplacing,positioning, arrows.meta}

\usepackage{listings}

\newcommand{\ImageWidth}{11cm}
\newcommand\blfootnote[1]{%
  \begingroup
  \renewcommand\thefootnote{}\footnote{#1}%
  \addtocounter{footnote}{-1}%
  \endgroup
}

\begin{document}

\title{Simulating Relational Event History Data: Why and How}
\author[1]{Rumana Lakdawala \thanks{\textbf{Corresponding Author:} : r.j.lakdawala@tilburguniversity.edu}}
\author[1]{Joris Mulder}
\author[2,3]{Roger Leenders}

\affil[1]{Department of Methodology and Statistics, Tilburg School of Social and Behavioral Sciences, Tilburg University, The Netherlands}

\affil[2]{Jheronimus Academy of Data Science, The Netherlands}

\affil[3]{Department of Organization Studies, Tilburg School of Social and Behavioral Sciences, Tilburg University, The Netherlands}


\date{}

\maketitle

\blfootnote{\begin{minipage}{\textwidth}
{Author profiles:} \\
\textbf{Rumana Lakdawala:} \url{https://scholar.google.com/citations?hl=en&user=x4z23RAAAAAJ}  \\
\textbf{Joris Mulder:} \url{https://scholar.google.com/citations?hl=en&user=W8EG7_gAAAAJ} \\
\textbf{Roger Leenders:} \url{https://scholar.google.com/citations?hl=en&user=63GISngAAAAJ}
\end{minipage} }

\begin{abstract}
    Many important social phenomena are characterized by repeated interactions among individuals over time such as email exchanges in an organization or face-to-face interactions in a classroom. To understand the underlying mechanisms of social interaction dynamics, statistical simulation techniques of longitudinal network data on a fine temporal granularity are crucially important. This paper makes two contributions to the field. First, we present statistical frameworks to simulate relational event networks under dyadic and actor-oriented relational event models which are implemented in a new R package \texttt{remulate}. Second, we explain how the simulation framework can be used to address challenging problems in temporal social network analysis, such as model fit assessment, theory building, network intervention planning, making predictions, understanding the impact of network structures, to name a few. This is shown in three extensive case studies. In the first study, it is elaborated why simulation-based techniques are crucial for relational event model assessment which is illustrated for a network of criminal gangs. In the second study, it is shown how simulation techniques are important when building and extending theories about social phenomena which is illustrated via optimal distinctiveness theory. In the third study, we demonstrate how simulation techniques contribute to a better understanding of the longevity and the potential effect sizes of network interventions. Through these case studies and software, researchers will be able to better understand social interaction dynamics using relational event data from real-life networks.
\end{abstract}

\noindent\rule{\textwidth}{0.5pt}


\newpage
\section{Introduction}
\label{sec1}
Our understanding of social interaction mechanisms between actors has been enhanced by dynamic network-based approaches. These actors can represent individuals, groups of individuals, organizations, or even countries. A network approach allows for the representation of actors by vertices and an event or interaction between the actors by the edges between the vertices. In traditional, static network analysis, edges are assumed to be stable and their presence or absence is indicative for the state the network is in. In continuous time dynamic network analysis (which is the focus of this paper), edges can appear and/or dissolve in real time at any time. To distinguish between edges that are stable and edges that can be in constant flux, the latter are often termed ``relational events."
The analysis of relational event data has the potential for a deeper understanding of the underlying social structures and how these evolve over time. This allows for the investigation of relationships between actors, dyads, and collective behaviour in the network. Particularly the temporal information yields insight into how variation in individual level dynamics affects network level characteristics, which in turn further affect individual behaviour.
Several statistical methods have been developed to study dynamic relational event history data for longitudinal social network analysis \citep{holland_1977_DynamicModelSocial,snijders_2010_IntroductionStochasticActorbased, hanneke_2010_DiscreteTemporalModels,krivitsky_2014_SeparableModelDynamic,perry_2013_PointProcessModelling}. Of particular interest to us is the relational event models proposed by \cite{butts_2008_RelationalEventFramework} and \cite{stadtfeld_2014_EventsSocialNetworks}, which allow dynamic social network analyses when the exact time-stamp of the events are available. In these relational event models, the event rate between actors is modeled as a function of the actors' past history and exogenous information available about the actors (nodal covariates) or pairs of actors (dyadic covariates) in the network. These models provide a flexible framework for network inference that allows for estimation of various drivers of social interaction behavior, such as homophily, reciprocity, transitivity, and preferential attachment. 

Despite its usefulness to quantify the relative importance of network drivers of the social evolution, truly understanding the underlying social mechanisms between actors based on relational event history data remains an enormous challenge. For example, it may be that different relational event models may induce similar social interaction behavior or the true underlying social mechanisms may be too complex to discover using a relational event analysis only. This is echoed in \cite{box_1976_ScienceStatistics} famous quote that ``all models are wrong...''. Simulation techniques have the potential to tackle these important current challenges in temporal social network modeling due to their ability to assess the impact of model specifications on data characteristics in a direct manner.

For this reason, the aims of the current paper is two fold. First it further develops and extends simulation techniques under flexible relational event modeling frameworks. 
Simulation frameworks are proposed under dyadic relational event models \citep{butts_2008_RelationalEventFramework} and under actor-oriented relational event models \citep{stadtfeld_2017_InteractionsActorsTime}. Unlike existing approaches, the proposed simulators support time-varying network effects, constrained risk sets, and various memory decay function to compute endogenous statistics allowing the study of the elapsed time of past events on the event rate in the (near) future. Second, the paper explores the potential of simulation techniques to improve social network analyses and to better understand social interaction behavior in real-life applications. Specifically, relational event simulators can contribute to the following objectives: 
\begin{enumerate}[label=(\roman*)]
    
    \item Assess goodness of fit: For a richer goodness of fit assessment, key network characteristics of the empirical relational event networks (such as degree distributions, density, triadic sub-structures, inter-event time distribution) can be compared to the simulated networks under the fitted model.
    \item Evaluate network interventions: Relational event simulations can be used to evaluate network interventions, by investigating the temporal dynamics of social networks under various intervention scenarios. For instance, researchers can use simulations to explore how quickly networks respond to interventions, how long interventions need to be carried out to achieve desired outcomes, and whether the effects of interventions persist over time or fade away. Additionally, simulations can be used to study the effects of targeting specific actors in the network, and to identify the most effective intervention strategies for different network configurations.
    
    \item Develop theory: Relational event simulations can be a powerful tool to develop social theories by providing a way to test and refine theoretical models in a controlled and systematic way. Simulations can be used to validate, test and extend existing theories as well as developing new theories about the social phenomena that drive the interactions in social networks. Moreover, simulations can help identify gaps in existing theories or data. For instance, if a simulated network does not match observed network characteristics, this could indicate that the theoretical model is incomplete or that data collection methods need to be revised. By identifying these gaps, researchers can develop more comprehensive theories and improve data collection methods.        
    \item Analyze the impact of network characteristics: By simulating networks with different characteristics (i.e. network size, combinations of endogenous and (or) exogenous effects, interaction effects, scaling of covariates, etc.) researchers can gain insight into how intricate combinations of the network settings affect social network dynamics. 
    \item Make predictions: Relational event simulations fitted with predictive models can help us understand how relational sequences may evolve beyond the observation period, enabling us to make predictions about the most likely events in the near future or explore alternate scenarios for how network dynamics may develop over a longer period.
    In the latter case, it is worth noting the well-known statistical issue of the Hamill forecast \citep{hamill_2001_InterpretationRankHistograms}, where a forecast is "wrong" in specific instances but "correct on average". When a researcher marginally calibrates the model, each prediction of a collection of events may be unrealistic, but the model can still accurately predict the occurrence of events on average \citep{gneiting_2007_ProbabilisticForecastsCalibration}. 
    
    \item Evaluate model sensitivity and statistical power: Simulation-based analysis can be used to assess the sensitivity of relational event models to specific model parameters and violations of assumptions such as proportional hazards and conditional independence. Researchers can also perform extensive analysis of power, accuracy, and precision of relational event models \cite[e.g.,][]{schecter_2020_PowerAccuracyPrecision}  under idealized circumstances to test and benchmark novel extensions of the model where the truth conditions that generated the data are known.
\end{enumerate}
To keep the paper as concise as possible, the paper shall focus on the first three objectives: goodness of fit assessment of relational event models, theory building on social interaction dynamics, and the planning network interventions.

This article is organized as follows. In Section \ref{sec2}, we introduce the dyadic and actor-oriented relational event models and outline the general-purpose simulation frameworks for these models that can be customized in various ways using problem-specific information. Next we illustrate a range of simulation scenarios that showcase the broad potential of the simulation frameworks in an attempt to inspire and guide new research. In Section \ref{sec-gof} we demonstrate how the simulations can be used to assess goodness of fit of relational event models using an email network from an organization. In Section \ref{sec-tb}, we describe how simulations can be used to develop theories about social phenomena that drive interactions and provide an example by simulating a well known social theory and test a boundary condition on this theory using simulations. In Section \ref{sec-intv} we discuss how the simulation framework can be used to investigate outcomes of network-based interventions. We demonstrate how strategies for network-based interventions can be simulated and compared on an organizational network.   Finally, we conclude with a discussion on limitations and future prospects of the simulation techniques in Section \ref{sec6}.

\section{Simulating Relational Event Networks}\label{sec2}
A relational event can be thought of as an event in which a sending actor (e.g. a person, group of individuals or other entity) directs an action to a receiving actor (e.g. another person, group, organisation, etc.) in the form of a discrete instantaneous event at a certain time point. The specification for an observed event $e = (i,j,t) $ entails a sender $i \in \mathcal{A}$, a receiver $j \in \mathcal{A}$, and $t$, the time at which the event was observed, where $\mathcal{A}$ is a set of actors \footnote{Without loss of generality, we keep the set $\mathcal{A}$ fixed over time in this paper}. An event represents one step in the dynamic network, and a temporal sequence of relational events forms the relational event history \citep{butts_2017_RelationalEventApproach}. In this paper, we describe networks in which the ties are directed (i.e an event from $i \rightarrow j$ is distinct from $j \rightarrow i$) and do not contain self loops (i.e from $i \rightarrow i$). However, extensions to undirected ties and networks with self-loops are straightforward. 

In order to simulate a relational event network sequence\footnote{Relational Event Network, Relational Event Histories, and Relational Event Sequence are referred to analogously.} $E$ in a time window $[0, \tau)$, every event $e$ in the sequence $E$ should be specified. This involves generating event times $ t \in [0,\tau)$ at which each event occurs and the
dyad $(i,j)$ associated with each event. This process of generating event sequences can be described under the dyadic Relational Event Model (REM) of \cite{butts_2008_RelationalEventFramework} as well as the actor-oriented Dynamic Network Actor Model (DyNAM) of \cite{stadtfeld_2017_InteractionsActorsTime}. We describe each of these simulation methods in detail in the following sections.

\subsection{Simulation Framework 1: Relational Event Model (REM)}
Consider a sequence of $M$ events $E = \{ e_1, e_2, \dots e_M\}$ such that. $t_m < t_{m+1}$. An event $e_m = (i_m,j_m,t_m) $, entails a sender $i_m \in \mathcal{A}$, a receiver $j_m \in \mathcal{A}$, and the time at which the event was observed, $t_m$. The set of possible events that can occur at any time is referred to as the risk set $\mathcal{R} \subseteq \{ (i,j): i,j \in \mathcal{A}\}$. The risk set may be constrained such that certain actors cannot interact with one another. Therefore the constrained riskset can be fixed or vary exogenously, where the possibility of occurrence of certain events is determined by exogenous factors such as availability of actors, location, or other circumstances specific to the research setting.

In REM, each dyad has its own rate of occurrence. A higher rate implies that the event involving that dyad is more likely to occur soon, and a lower rate implies that the occurrence of the event is more rare. The rate is modeled as a log linear function of the endogenous and exogenous statistics pertaining to that dyad along with the parameters $\boldsymbol{\beta}(t)$ that represent the strength of these statistics to explain social interaction behavior in the network. The occurrence of events is modelled using a piece-wise constant hazard model. Under the piece-wise constant model, the rates are assumed to only change when an event occurs (anywhere in the network). Thus, as time progresses and events are observed, the rate is updated to reflect the new network history. The rate $\lambda^{dyadic}_{ij}$ of a dyad $(i,j)$ at time $t$ is then specified in a log linear form as:

\begin{equation} \label{eqn:lambda-tie}
\lambda^{dyad}_{ij}(t ) = \begin{cases} 
    \text{exp}\{ \boldsymbol{\beta}(t)^T \; X(i,j,E_{t}) \} & \  (i,j) \in \mathcal{R} \\
    0 & \ (i,j) \notin \mathcal{R}
\end{cases}
\end{equation}

where $X(i,j,E_{t})$ is a vector of $P$ statistics for the dyad $(i,j)$ on the event sequence $E_t$ until time t  and $\boldsymbol{\beta}(t) \in {\rm I\!R^P}$ is the vector of corresponding parameters at time $t$ (the parameters may be constant and not-time varying in which case, $\beta$ is a constant. The statistics vector  $X(i,j,E_{t})$ can capture either endogenous network characteristics as a function of the past events in the history $E_{t}$ before time $t$ or exogenous actor or dyadic attributes. The notation for rate has a superscript `dyad' to distinguish it from the rate under the actor-oriented model introduced in the next section.

Under the piece-wise constant hazard model, the waiting time $\delta_m = t_{m} - t_{m-1}$ between subsequent events are assumed to be conditionally independent and are specified by an exponential distribution, i.e.,
\begin{equation}\label{eqn:waiting-time-tie}
p^{dyad}(\delta_m|E_{t_{m-1}},\boldsymbol{\beta}(t_{m-1}),X) \sim \text{Exp}(\Lambda^{dyad}(t_{m-1} ))
\end{equation}
with the cumulative rate as distributional parameter, given by
\[
\Lambda^{dyad}(t_{m-1})  = \sum \limits_{(i,j) \in \mathcal{R} } \lambda_{ij}^{dyad}(t_{m-1}).
\]
The probability of a dyad $(i,j) \in \mathcal{R} $ to be involved in the next observed event, follows a multinomial distribution where the probabilities are proportional to the dyadic rates: 
\begin{equation}\label{eqn:prob-tie}
    \begin{split}
    p^{dyad}(i,j \mid t , \boldsymbol{\beta}(t), X, E_{t}) = 
    \frac{  \lambda^{dyad}_{ij}(t ) }{ \Lambda^{dyad}(t )  }.  \;
    \end{split}
\end{equation}

In order to simulate relational event networks, the user needs to specify which endogenous or exogenous network statistics \citep[e.g.,][]{leenders_2016_OnceTimeUnderstanding} are included in the model and the magnitude of the corresponding parameters (based on theory, a specific dynamic of theoretical interest, or a fitted model). If preferred, it also possible to initialize simulations with a pre-defined starting sequence $E_0$. This would be useful if a user wants to predict the immediate future of an empirical network or to simulate using different models from the same event history. The inputs to the simulation algorithm play an important role in determining the characteristics of the sequence output. The number of actors $N$ in the network determines the size of the riskset as $N(N-1)$ (when all dyads are at risk) and $\tau$,  the time until which the simulation algorithm is run, determines the number of events in the sequence. The parameters $\boldsymbol{\beta}(t)$ capture the sign and magnitude of the effect of the statistics. Given these inputs, a relational sequence can be simulated 
using Algorithm \ref{algo:sim-rem}.

\begin{algorithm}[ht]
    \DontPrintSemicolon
    \SetKwInOut{Input}{input}
    \SetKwInOut{Output}{output}
    \Input{$\boldsymbol{\beta}, \tau , \mathcal{R}$, $E_o$}
    \Output{sequence $E$}
    \eIf{$E_o$ is empty}{Initialize event sequence $E$ with a random event $e_1$ \;}{Initialize event sequence $E = E_o$\;}
    \While{$t_m \leq \tau$}{
        compute statistics matrix X on $E_{t_{m-1}}$\;
        
        update event rates $\lambda^{dyad}_{ij}(t_{m-1} \mid E_{t_{m-1}},\boldsymbol{\beta}(t), X) \ \forall (i,j) \in \mathcal{R}$ using Equation \eqref{eqn:lambda-tie}\;
        
        sample a dyad $(i_m,j_m) \in \mathcal{R}$ from multinomial distribution in Equation \eqref{eqn:prob-tie} \;
        
        sample inter-event time $ \delta_m \sim \text{Exp} \left( \sum \limits_{(i,j) \in \mathcal{R} } \lambda^{dyad}_{ij}(t_{m-1} ) \right)$  \;
        
        $t_m = t_{m-1} + \delta_m$ \;
        
        append $(i_m,j_m,t_{m})$ to $E$ \;
        update $\boldsymbol{\beta}(t)$ if time-varying \;
        $m++$\;
        }
    
    \caption{Simulation of a relational event sequence under dyadic model}
    \label{algo:sim-rem}
\end{algorithm}

\subsection{Simulation Framework 2: Dynamic Network Actor Model (DyNAM)}
\label{sec2:dynam}
Instead of all dyads competing with each other to participate in the next event, under the DyNAM the actors compete with each other to becomes the next sender, where each actor has its own rate parameter to be the next sender. A higher rate implies that the actor is more likely to be the sender of the next event. The rate of the actor as a sender is specified similarly to the rate parameter in a dyadic REM using a log linear model of endogenous and exogenous sender statistics, which are now summarized in the statistics matrix $X^s$. The time of the next event then follows an exponential distribution,
\[
p^{actor}(\delta_m|E_{t},\boldsymbol{\gamma}(t),X^s)\sim \text{Exp}(\Lambda^{sender}(t)),
\]
where $\Lambda^{sender}(t)  = \sum \limits_{i \in A } \lambda_{i}^{sender}(t)$, with $\mathcal{R}^s$ denotes the set of actors that are at risk to become a sender, and the rate parameter of actor $i$ to become a sender is defined by
\begin{equation} \label{eqn:lambda-actor}
     \lambda_{i}^{sender}(t) = \exp\{
\boldsymbol{\gamma}(t)^TX^s(i,E_t),
\}
\end{equation}

where $\gamma$ denotes the vector of coefficients which quantify the importance of the respective statistic in $X^s$.
Furthermore, the probability of actor $i$ to become a sender is proportional to the respective rate parameters according to a multinomial distribution

\begin{equation} \label{eqn:prob-sender-actor}
    p^{actor}(i \mid E_{t} , \boldsymbol{\gamma}(t),X^s) = \frac{\lambda_{i}^{sender}(t) }{ \Lambda^{sender}(t).  }
\end{equation}
Finally, the probability of actor $j$ to become the receiver is modeled conditionally on the sender $i$ using a multinomial distribution with probability
\begin{equation} \label{eqn:prob-reciever-choice}
    p^{actor}(j | i , t, E_{t} , \boldsymbol{\alpha}(t),X^r ) = \frac{\lambda_{j|i}^{receiver}(t) }{ 
    \Lambda_{i}^{receiver}(t)
     } 
\end{equation}
where the preference parameter of actor $j$ to be come the receiver is given by
\[
\lambda_{j|i}^{receiver}(t)=
\text{exp}(\boldsymbol{\alpha}(t)^T X^r(i,j,E_{t} ),
\]
which is modeled in a similar manner as the rate parameter,
and the summation is given by
\[
\Lambda_i^{receiver}(t)=
\sum_{j \in A \backslash \{i\}} \lambda_{j|i}^{receiver}(t).
\]

The simulation framework of relational event sequences under the DyNAM is then summarized in Algorithm \ref{algo:sim-dynam}

\begin{algorithm}[h]
    \DontPrintSemicolon
    \SetKwInOut{Input}{input}
    \SetKwInOut{Output}{output}
    
    \Input{$\boldsymbol{\gamma}, \boldsymbol{\alpha},\tau ,\mathcal{R},E_o $}
    \Output{sequence $E$} 
    \eIf{$E_o$ is empty}{Initialize event sequence $E$ with a random event $e_1$ \;}{Initialize event sequence $E = E_o$\;}
    \While{$t_m \leq \tau$}{
        compute sender statistics matrices $X^s(E_{t_{m-1}})$ \;
        compute sender rates $\lambda_{i}(t_{m-1} \mid E_{t_{m-1}} , \boldsymbol{\gamma}(t),X^s) \ \forall i \in \mathcal{A}$ using  Equation \eqref{eqn:lambda-actor}\;
        sample inter-event time $ \delta \sim \text{Exp} \left( \sum \limits_{(i) \in \mathcal{A} } \lambda_{i}(t_{m-1} \mid E_{t_{m-1}} , \gamma,X^s) \right)$  \;
        sample a sender $i_m \in \mathcal{A}$  from categorical distribution in Equation \eqref{eqn:prob-sender-actor}\;
        compute receiver statistics matrix $X^r$ \;
        sample a receiver $j_m \in A \backslash \{i_m\}$ given sender $i_m$ from categorical distribution in   \eqref{eqn:prob-reciever-choice}\;
        append $e_i = (i_m, j_m, t_m)$ to $G$ \;
        update $\boldsymbol{\gamma}(t)$ and $\boldsymbol{\alpha}(t)$ if time-varying \;
        $m++$\;
    }
    \caption{Simulation of relational event sequence under actor-oriented model}
    \label{algo:sim-dynam}
\end{algorithm}

The main differences between the dyadic relational event model and its actor-oriented counterpart (i.e., the DyNAM) can be understood from how they model the occurrence of relational events. In the dyadic REM, the timing of the next event and the dyad that is observed are modeled using dyad specific rate parameters. Alternatively, in the actor-oriented model, the timing and the sender of the next event are modeled via actor (sender) specific rate parameters, and the next receiver is modeled conditionally on the sender (i.e., \textit{chosen} by the sender) where all actors have separate rate parameters as potential receivers  \citep{stadtfeld_2017_InteractionsActorsTime}. These differences in modelling are also reflected in the simulation frameworks. In the dyadic model, all dyads are competing to be sampled as the next observed event whereas in actor-oriented model, the actors first compete to become the sender and then the remaining actors compete to become the receiver that is chosen by the given sender. The two models also differ in how they conceive network change. In the dyadic model, the building block is the dyad and the model assumes the two participating actors in the dyad to actively determine whether and when they will interact. In the actor-oriented model, on the other hand, the agency is at the level of the sender, who determines when to get active. Therefore, the latter model is conceptually actor-driven rather than dyad-driven. In the simulation frameworks this difference translates to which statistics are used to determine the next sampled event. In case of the dyadic REM, dyadic statistics are used to sample the next event whereas in case of actor-oriented model, actor-focused statistics are used to sample the sender of the next event and dyadic statistics are used to make the choice of the receiver given the sender.

\subsection{\texttt{remulate}: R package to simulate relational event histories}
This article introduces the R package \texttt{remulate} \citep{remulate}. The package is developed to assist researchers in simulating relational event histories. The package aims to reduce the complexity associated with simulation techniques and to make them more accessible to social network researchers. The package enables the user to simulate using a wide range of commonly used exogenous and endogenous statistics for the tie- and actor-oriented relational event model approaches as well as several important extensions.

Unlike \texttt{relevent} \citep{relevent} which also allows users to simulate relational event histories, \texttt{remulate} has the following additional features:
\begin{enumerate}
\item Support for dyadic relational events models as well as actor-orientied relational event models.
\item Support of a larger collection of pre-computed endogenous statistics with various methods of normalization and standardization and support for interaction terms.
\item Support different types of memory decay functions, such as exponential decays or step-wise functions, to study the influence of the transpired time of past events on the event rate between actors in the future.
\item Support for constrained risk sets, such that if two actors cannot interact, events involving that dyad would not be sampled.
\item Support time-varying network parameters to study  instantaneous, gradual, and periodic changes of drivers of social interaction behavior.
\item Support for frailty relational event models to capture nodal heterogeneity.
\item Support for relational event block models.
\end{enumerate}

Appendix \ref{appendix:A} presents exemplary R code on how to simulate from tie-oriented and actor-oriented relational event models using the \texttt{remulate} package. 

\section{Assessing goodness-of-fit through simulations}\label{sec-gof}
 In the following we will present extensive applications that highlight why relational event simulators are essential for temporal social network research. We begin with assessing goodness of model fit through simulations. 
 
Typical approaches to evaluating goodness-of-fit of relational event models involves a comparison of different models by balancing parsimony and accuracy. The preferred model generally contains the fewest parameters while yielding satisfactory fit for the observed relational events. This usually involves using information criteria such as the Akaike Information Criterion (AIC), the Bayesian Information Criterion (BIC), or other likelihood-based goodness-of-fit measures. Although these criteria are useful as a relative measure of fit when comparing models, they do not give an indication of the model fitness in an absolute sense neither do they provide a substantive direction of possible misfit. Thus, when comparing multiple relational event models using such general information criteria, the ``best'' model may still have a very poor fit to the observed event sequence in terms of capturing important network characteristics and it may result in poor predictions.

On the other hand, simulation-based methods can be used to assess model fit in an absolute sense by assessing whether important network characteristics in the data are also present in the simulated data using the fitted model. If simulated sequences based on a fitted model bear little resemblance to the observed sequence (or misses important characteristics of the network dynamic), this generally suggests a poor fit of the model to the data. 
Misfit can occur due to inclusion of effects that are not operative in the data or due to exclusion of key effects or external factors that play an important role in the network dynamics. Inferences based on misfitted models are therefore (theoretically) unreliable, and simulation-based methods are needed to assess whether the model results in a reasonable fit to the observed data sequence before drawing inferences.

Before describing the general methodology, we briefly mention some previous work on Simulation-based methods for model assessment in social network research. \cite{hunter_2008_GoodnessFitSocial} introduced simulation-based goodness-of-fit tests for social networks under exponential random graph models (ERGMs) using structural indices such as degree distributions, edgewise shared partners, and geodesic distance distributions.  \cite{snijders_2015_RepresentingMicroMacro} introduce a variety of structural fit indices to describe cross sectional data such as the size of the largest component, number of components, median geodesic distance, transitivity coefficient, variance of in or out degree divided by mean, correlation between in and out degrees, graph hierarchy and, least upperboundedness. \cite{lospinoso_2019_GoodnessFitStochastic} include higher order indices in the form of triad census to assess if the simulated networks accurately resemble triadic structures in the observed network. \cite{wang_2020_ModelAdequacyChecking} describe behavioral indices to assess fit for models that jointly model behaviour and network structure. \cite{brandenberger_2019_PredictingNetworkEvents} considered the accuracy of event predictions to assess the goodness-of-fit of relational event models for political networks. Because relational event models not only model the network structure but the timing and order of events as well, goodness-of-fit tests would be incomplete without assessing the fit of candidate models on temporal indices. It is possible that the model fits data well within certain time periods but not in others or that events between certain dyads in the predicted dataset are further apart than in the observed network. In addition, it is possible to compute structural indices that respect the timing and order of events in relational event datasets, such as temporal degree, temporal reachability, and temporal betweenness \citep{falzon_2018_EmbeddingTimePositions}. \cite{nicosia_2013_GraphMetricsTemporal} similarly define temporal connectedness, various temporal centrality indices and temporal shortest paths for time-respecting paths in dynamic networks. These temporal indices enhance the fit indices used for static or longitudinal networks, by also incorporating the timing information available from relational event datasets. In addition \cite{amati_2024_GoodnessFitFramework} suggest examining internal time structures in relational event models to capture the specific sequence and timing of events, such as reciprocity and transitive closure to ensure that the model accurately reflects the temporal patterns observed in social interactions.

The indices mentioned above are just a few examples from the social network literature to assess the goodness of fit. In fact the number of temporal network characteristic to assess is unlimited. Moreover when assessing goodness of fit, we believe that researchers should focus on reproducing specific aspects of the data 
that are of theoretical and/or practical interest rather than applying a default set of statistics.  Ultimately, the choice of the fit indices should depend on the specific phenomena the researchers intend to address in their analysis. Therefore, it is impossible to provide an exhaustive set of network characteristics for general use in relational event analysis. Instead, we provide a general step-by-step methodology to assess the goodness of fit in relational event modeling applications. Subsequently, we apply the methodology to a real-life case study.

\subsection{General methodology}
\label{sec:gof-alg}
 Suppose the goal is to assess the goodness-of-fit of a candidate model $M_c$ for the observed sequence $\mathcal{E}$ on a set of fit indices $f(\mathcal{E})$. The fit indices can be a single value (in case the dynamic network is aggregated) or a vector of values measured during various time intervals across the observation period. The steps for evaluating goodness-of-fit using this approach are:

\begin{enumerate}
    \item Identify relevant indices $f$ of the observed sequence against which the goodness-of-fit will be evaluated.
    \item Compute the corresponding indices on the observed sequence, i.e. compute $f(\mathcal{E}^{obs})$ on $\mathcal{E}^{obs}$.
    \item Estimate the parameters $\hat{\boldsymbol{\beta}}_{M_c}$ of the candidate model $M_c$.
    \item Generate $L$ sequences $ E^{sim}= \{ \mathcal{E}_1, \mathcal{E}_2 \dots \mathcal{E}_L \}$ from the estimates $\hat{\boldsymbol{\beta}}_{M_c}$ using the simulation frameworks described in Section \ref{sec2}.
    \item Compute the fit indices $f(\mathcal{E}_l)$ for each of the simulated sequences $\mathcal{E}_l \in E^{sim}$. 
    \item The resulting distribution of $L$ goodness-of-fit indices $f(\mathcal{E}_1) , f(\mathcal{E}_2) \dots f(\mathcal{E}_L) $ can be compared with $f(\mathcal{E}_o)$ visually \citep{hunter_2008_GoodnessFitSocial} or quantitatively \citep{lospinoso_2019_GoodnessFitStochastic,chen_2019_BootstrapMethodGoodness} to evaluate the fit of $M_c$ on the observed data. If the computed fit indices (of their distributions) on the $L$ sequences differ greatly from $f(\mathcal{E}^{obs})$, it can be concluded that $M_c$ does not fit the data appropriately and further exploration is needed to determine where the model may be lacking.
\end{enumerate}

In the context of relational event models, this approach to evaluating goodness-of-fit (i) allows researchers to test the model's fit across a set of interpretable structural and social indices, and (ii) guides the selection of network effects to be included in the model by pointing the researcher towards a direction in which the model may be misfit or lacking in specification. 

This approach to evaluating the fit is very flexible in terms of the indices against which the fit can be evaluated. The indices can be global network properties such as network density, degree variances, or local properties such as reciprocated ties, or transitivity. They can be based on social behaviour, for example, homophily on specific attributes, subgroups in data, or proportion of actors in a network exhibiting behaviours such as smoking, alcohol use, et cetera, or can be purely structural such as clustering or connected components. The flexibility also allows researchers to monitor these indices across all time points or networks aggregated over sub-intervals of the observation period.

\subsection{Example: Assessing goodness-of-fit}
\label{sec:ex-gof}
To demonstrate the evaluation of goodness-of-fit through simulations, we evaluate the fit of a model presented in a recent study by \cite{gravel_2023_RivalriesReputationRetaliation}. The study focuses on the dynamics of inter-gang violence in Los Angeles using a relational event modeling approach. The relational event history comprises of violent incidents involving 33 gangs spanning over three years. The authors examine various mechanisms that drive the continuation of violence in this network. The authors focus on how gang characteristics, spatial proximity, enduring rivalries, and how the dynamics of past conflicts influence future violent events. The authors argue that while retaliation is a significant mechanism, other factors like established rivalries and inertia play crucial roles in driving future violence. The authors of the article attempt to assess the goodness of fit of their model by checking how well the fitted model can predict the observed events in-sample: For every observed event, it was checked how likely it was to be observed given the (endogenous) predictor variables at that point. Even though this analysis indicated that the fitted model has a high predictive accuracy, this assessment is unable to answer whether the entire sequence of relational events is plausible under the fitted model when it would be used to generate entire event sequences ``from scratch''. That is the purpose of the current simulation analysis. Moreover, following step 1 of the goodness of fit assessment procedure presented in Section \ref{sec:gof-alg}, we shall investigate the fit of various data characteristics on gang violence including gang size, inter-dyadic distance, geodesic hop-pah lengths, degree statistics, and inter-event times between rival gangs.

\renewcommand{\arraystretch}{1}

\begin{figure}[htp]
     \centering
     \begin{subfigure}[b]{0.4\textwidth}
         \centering
    \includegraphics[width=\textwidth]{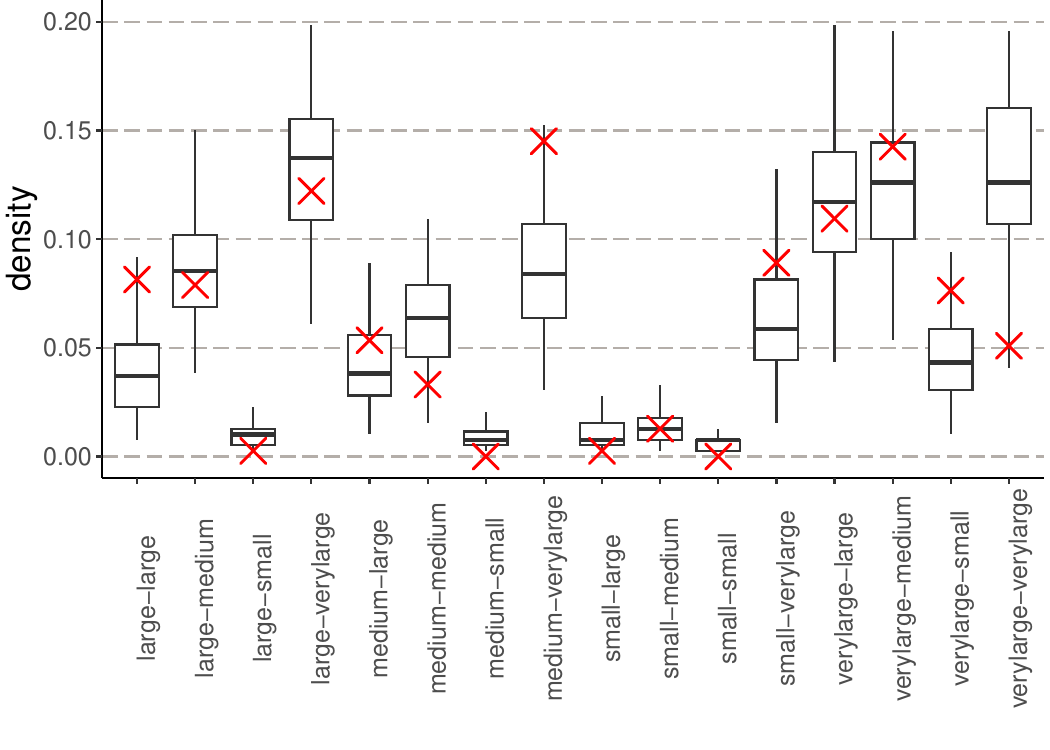}
         \caption{Inter-gang size event density}
         \label{fig:gofa}
     \end{subfigure}
     \begin{subfigure}[b]{0.4\textwidth}
         \centering
         \includegraphics[width=\textwidth]{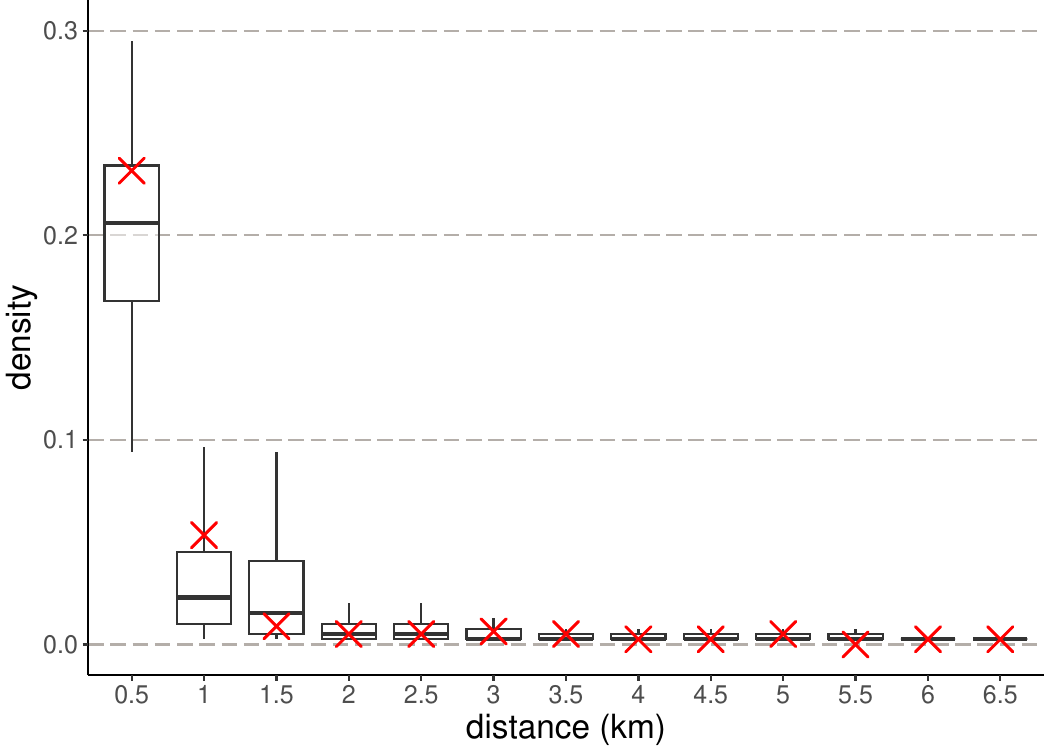}
         \caption{Inter-dyadic distance event density}
         \label{fig:gofb}
     \end{subfigure}
     \vfill 
     \begin{subfigure}[b]{0.4\textwidth}
         \centering
         \includegraphics[width=\textwidth]{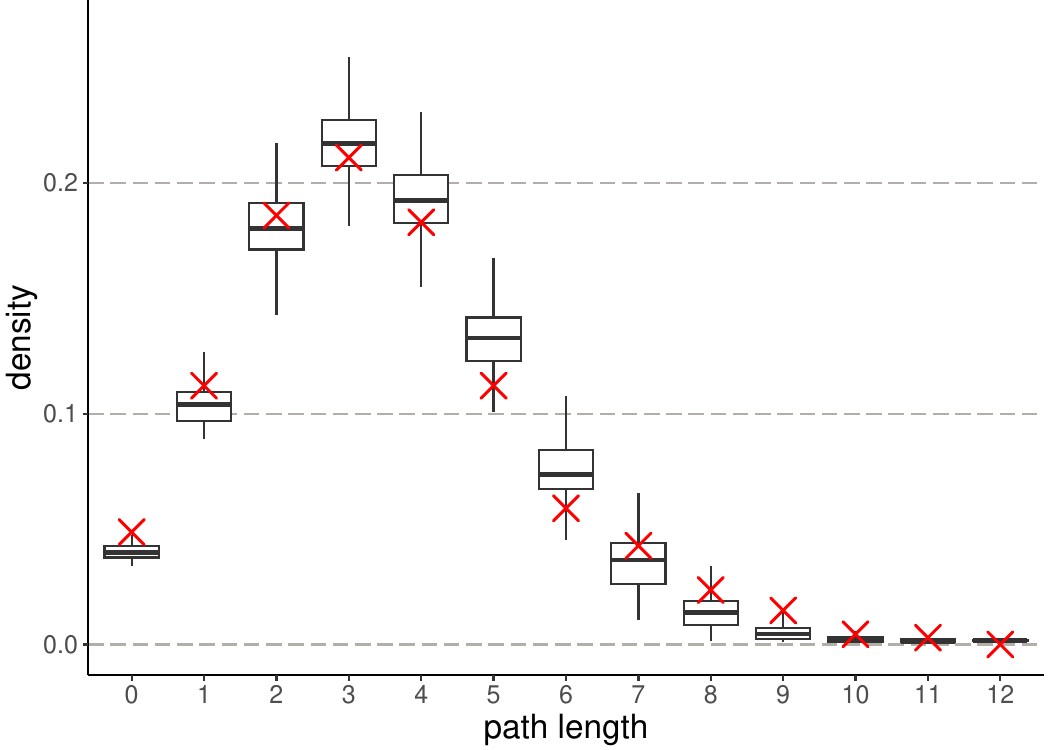}
         \caption{Temporal geodesic hop-path length distribution}
         \label{fig:gofc}
     \end{subfigure}
     \begin{subfigure}[b]{0.4\textwidth}
         \centering
         \includegraphics[width=\textwidth]{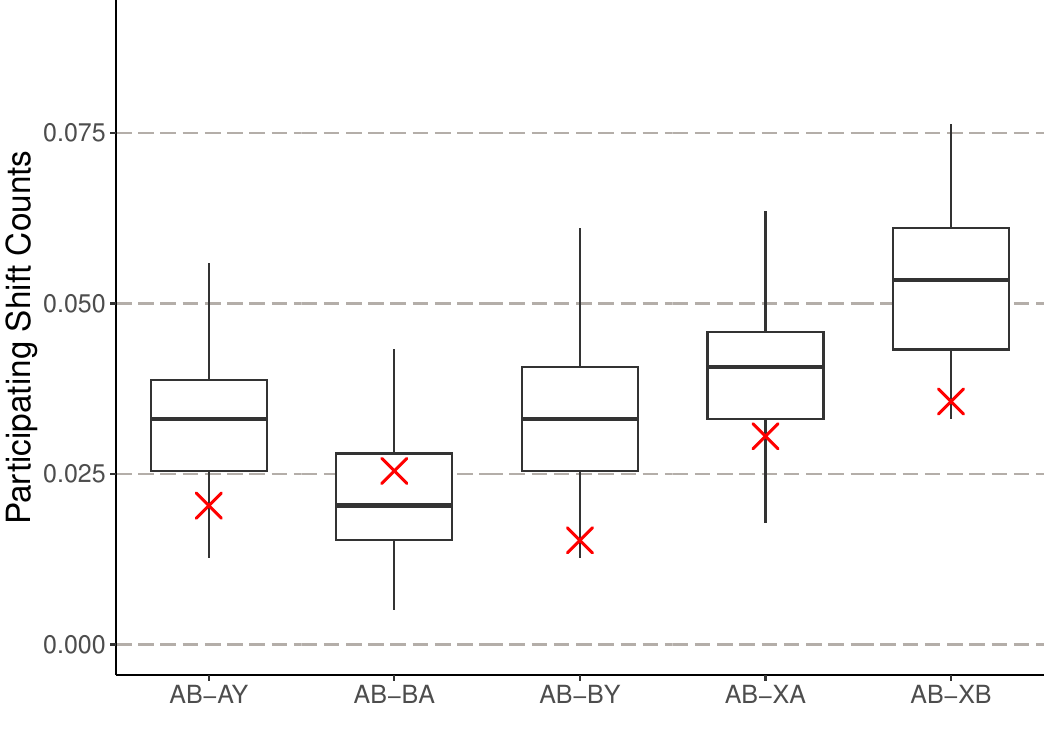}
         \caption{Participating shift counts }
         \label{fig:gofd}
     \end{subfigure}
     \vfill 
     \begin{subfigure}[b]{0.4\textwidth}
         \centering
         \includegraphics[width=\textwidth]{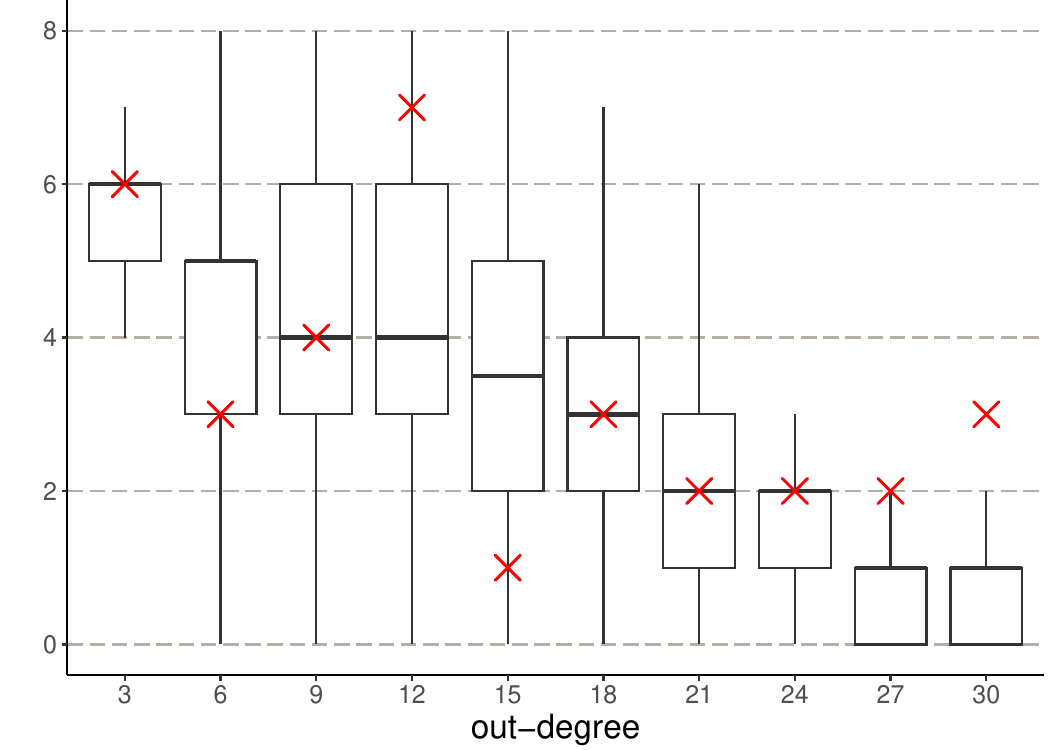}
         \caption{Out-degree distribution}
         \label{fig:gofe}
     \end{subfigure}
     \begin{subfigure}[b]{0.4\textwidth}
         \centering
         \includegraphics[width=\textwidth]{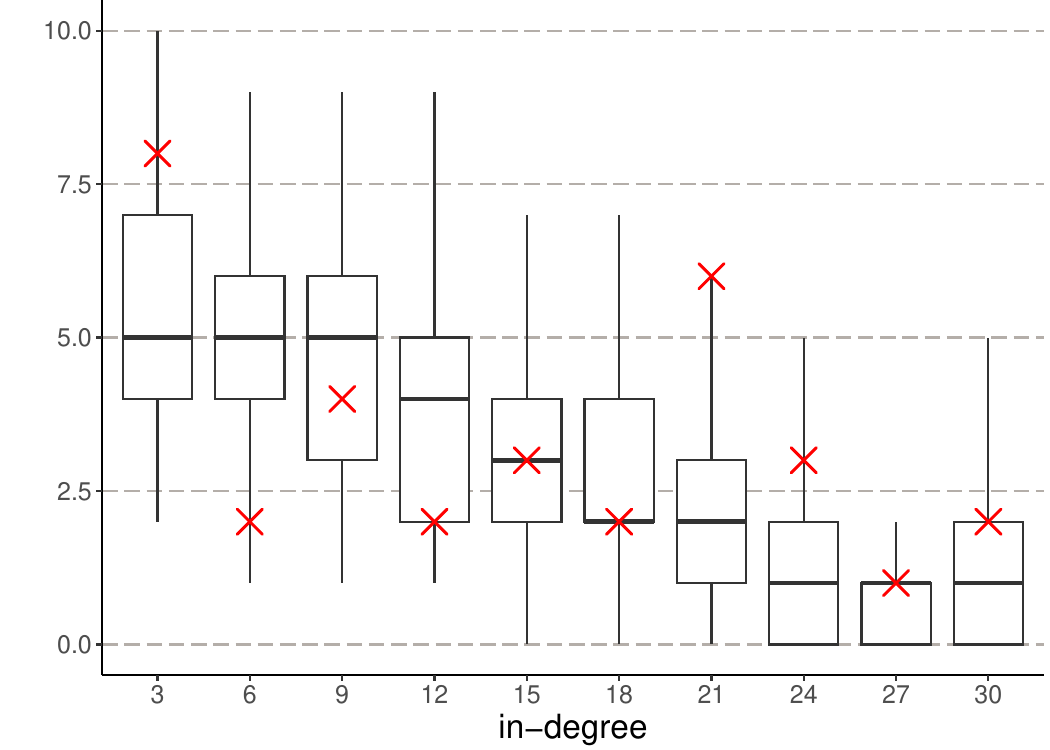}
         \caption{In-degree distribution}
         \label{fig:goff}
     \end{subfigure}
     \vfill 
     \begin{subfigure}[b]{0.4\textwidth}
         \centering
         \includegraphics[width=\textwidth]{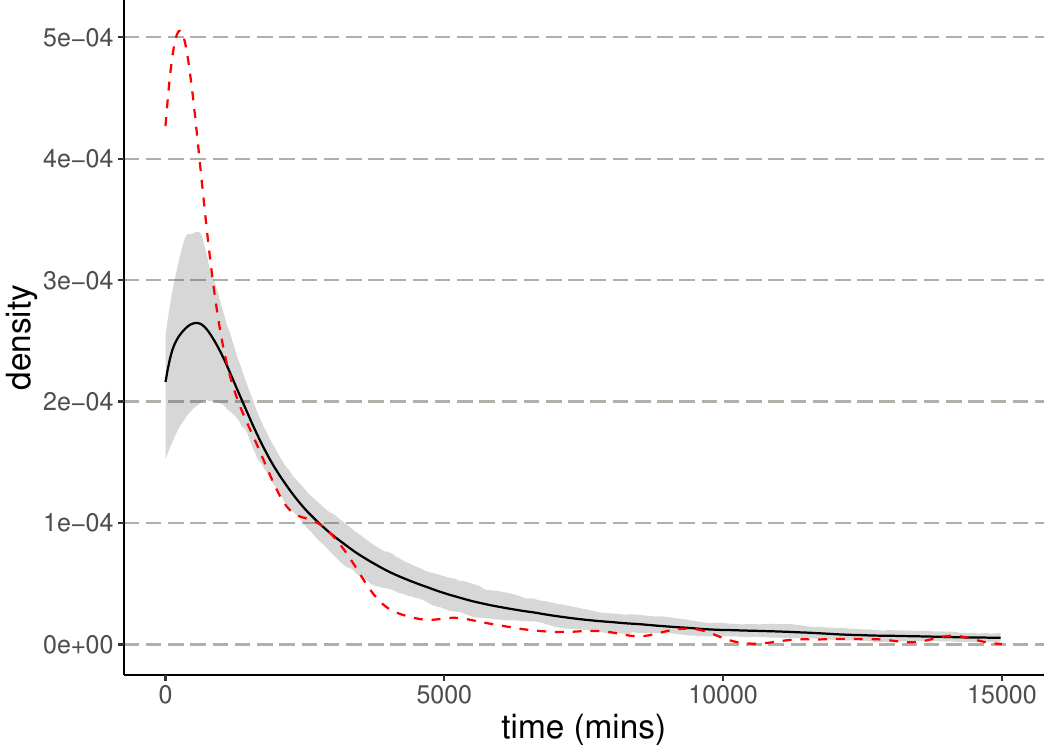}
         \caption{Interevent time - rivals}
         \label{fig:gofg}
     \end{subfigure}
     \begin{subfigure}[b]{0.4\textwidth}
         \centering
         \includegraphics[width=\textwidth]{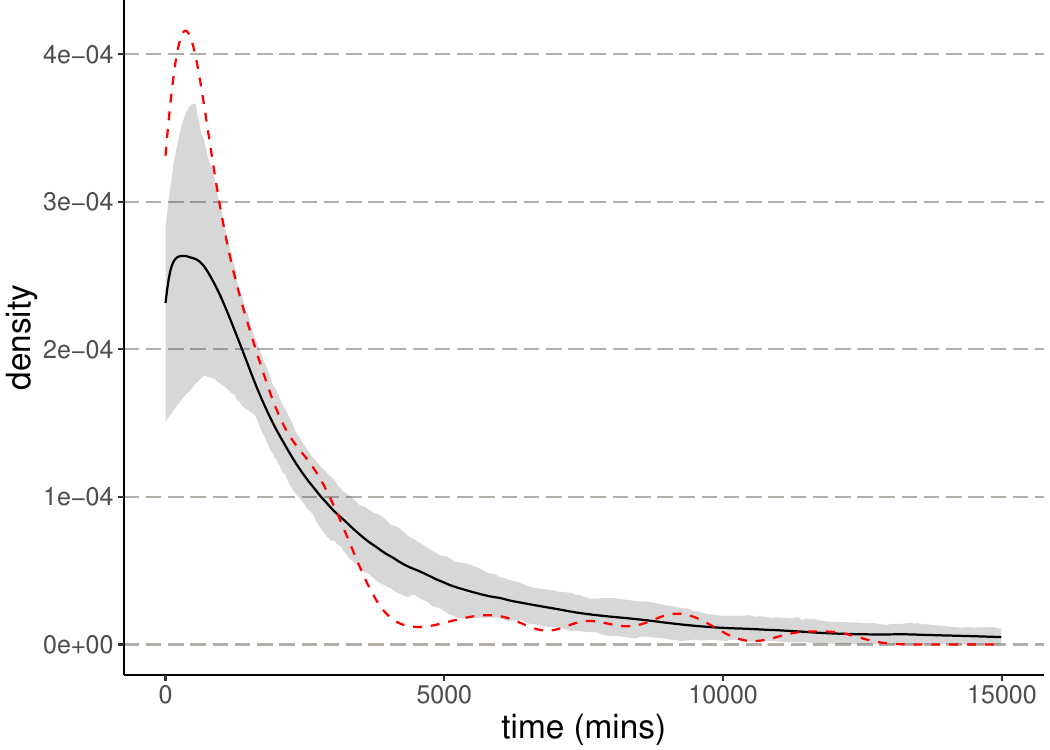}
         \caption{Interevent reciprocation time - rivals}
         \label{fig:gofh}
     \end{subfigure}
     \caption{Goodness-of-fit comparison. Each figure plots the values of the fit indices over 100 simulated sequences and the observed data. In figures (a-f) the red crosses represent the observed data and the boxplots depict the distribution of the fit indices across simulated networks. In figures (g-h) the dotted red line represents the observed data.}
        \label{fig:gof}
\end{figure}

To assess the goodness of fit for the relational event models presented in this study, we simulate from the fitted relational event models (see Appendix \ref{appendix:B} for R-code) using the maximum likelihood estimates presented in Table 2 of the study. The plots in Figure \ref{fig:gof} illustrate the goodness-of-fit for the model, comparing the characteristics of simulated networks against actual observed data in the context of street gang violence. Each subplot highlights a different network characteristic derived from violent conflicts between gangs. Figure \ref{fig:gof}(\subref{fig:gofa}) shows the density of events categorized by size of participating gangs in the dyads. In the context of the paper, understanding these interactions is crucial because larger gangs might have different resources, territorial ambitions, and levels of influence, which can affect their interaction patterns with smaller gangs. The observed data (red crosses) are relatively well-aligned with the median of the simulated distributions, implying that the model captures the frequency of events across gang-size effectively. There is a slight misfit involving dyads with a larger gang size which might necessitate a need for an additional variable to effectively reproduce the interaction patterns of larger sized gangs. In Figure \ref{fig:gof}(\subref{fig:gofb}) the density of events is plotted against the geographic distance between dyads. The article highlights the importance of spatial proximity in the dynamics of gang violence, making it vital for the model to accurately simulate these interactions. The model appears to accurately capture the decline in event density as distance increases. The model slightly underestimates the frequency of localized conflicts which are more frequent and significant within closer geographical proximity. The Figure \ref{fig:gof}(\subref{fig:gofc}) examines the distribution of temporal path lengths with shorter paths indicating more direct interactions over time between gangs (with lesser temporally respecting hops). The model reproduces the geodesic hop-path length distribution across the dyads in the network reasonably well. This fit index helps understand how violence spreads over time through the network, testing theories of violence contagion and diffusion within gang dynamics.

Participating shift counts of dyadic turn-taking events \citep{gibson_2003_ParticipationShiftsOrder} are depicted in \ref{fig:gof}(\subref{fig:gofd}). While, AB-BA and AB-BY type participating shifts were included in the model, the results suggest that the model is also able to reproduce the AB-AY, AB-XA and AB-XB turn taking events well. The out-degree distribution represents the number of conflicts initiated by each gang is depicted in \ref{fig:gof}(\subref{fig:gofe}). 
The observed data mostly aligns with the simulated medians for lower out-degrees but tends to fall outside the interquartile range for higher degrees. This could be due to the model not accounting for potential external factors that might influence a gang's propensity for aggressive conflict initiation. In addition we also look at the in-degree distribution of the data in Figure \ref{fig:gof}(\subref{fig:goff}) which suggest an adequate fit.

The last two figures \ref{fig:gof}(\subref{fig:gofg},\subref{fig:gofh}) show the distribution of interevent events categorized by inertia and reciprocation among rival gangs. These timing-based indices test the model's ability to quantify the rapidity and persistence of violence between rival gangs. It is vital for investigating how well the model replicates violence escalation and retaliation cycles, which are key aspects of gang dynamics described in the paper In the figure, the observed data, is represented by the red dotted line. Both plots demonstrate that conflicts between rival gangs are characterized by rapid sequences of events, with a significant number of actions occurring within a very short time frame following an initial event. Although, the plot for reciprocation times between rival gangs has a somewhat less steep decline compared to the interevent times for rivals. This suggests that while immediate responses are common, there's slightly more variation in the timing of reciprocations, possibly due to differing strategic considerations or external factors influencing the timing of a gang's response. The simulated results indicate an underestimation at very short inter-event times, suggesting the model might not fully capture the quickest escalations in gang conflict.  Refining the model by including an interaction term between the recent reciprocation and recent target persistence and rivalry might address the misfit. Additionally, short term-memory effects \citep{brandes_2009_NetworksEvolvingStep}
that allow for more rapid changes in event probabilities immediately following a conflict event might allow the model to address the misfit.

Evaluating the goodness-of-fit using simulations has provided us with a more nuanced understanding of how well the model fits and highlighted potential areas of misfit compared to predictive performance or likelihood based measures. Overall, while the model provides a good general fit to the observed data across several key indices, the detailed analysis provided insights into where the model falls short in capturing the dynamics of gang conflicts, such as the inter-event times between rival gangs. Understanding and refining these aspects of the model could lead to better predictions of gang conflict dynamics and more effective strategies for law enforcement and community interventions aimed at reducing the frequency and severity of these conflicts.

\section{Developing theories using simulations}\label{sec-tb}
The second important application of the relational event simulation techniques involves the development of social theories.
Many social phenomenon arise from repeated interactions among individuals in a social network over time. Undoubtedly, researchers continue to gain insights into the drivers of repeated social interactions through empirical studies. The relational event model itself has contributed greatly towards modelling the data arising out of such empirical research. While this is useful for estimating the magnitude of network effects or confirming differences between different groups in specific empirical settings, fitted REMs are typically based on theoretical ideas of the researcher and contribute to testing existing theories rather than being employed as a way to develop or refine (new) theory. Simulations can provide a powerful and flexible tool for (further) developing or expanding theories \citep{davis_2007_DevelopingTheorySimulation} regarding why, how, and when interactions occur in a social network. The proposed relational event simulation frameworks provide flexibility to combine multiple network effects such as transitivity, homophily, or preferential attachment in one theoretical model, allowing researchers to test and develop richer theories about social interaction dynamics. Below we first describe how we can (further) develop social network theories using relational event simulations, and subsequently, we present an extensive application of the methodology to better understand group formation using optimal distinctive theory.

\subsection{Building, evaluating, and extending social network theories}
Relational event simulations can be utilized to develop theories to a) study the emergence of social phenomena, b) evaluate boundary conditions of theories, and c) incorporate timing and dynamism in theories. We briefly elaborate on this below.

\textit{Study the emergence of social phenomena.} 
Emergence is a fundamental concept in social sciences, referring to the way collective phenomena arise from the interactions of individuals \citep{kozlowski_2000_MultilevelApproachTheory}. Social phenomena can be considered emergent when they cannot be reduced to the behavior of individuals, but instead emerge from complex interactions among them. Relational event sequences can help explain elements of the emergent properties that occur on a global level but are a consequence of the events, their sequence, timing and patterns associated with event occurrence.

Simulations offer a powerful tool for studying the emergence of social phenomena \citep{epstein_1996_GrowingArtificialSocieties,gilbert_2005_SimulationSocialScientist}, especially those that are unexpected or counter-intuitive. By simulating relational event models, researchers can generate scenarios that test the boundaries of existing theories and develop new insights into the mechanisms underlying emergent phenomena. In this sense, simulations offer a natural way of describing complex dynamics as a combination of familiar network mechanisms. Relational event simulation frameworks are particularly useful in studying emergent properties because they can capture the dynamic nature of social phenomena over time. These frameworks allow researchers to model how the behavior of individuals within a social network can lead to collective outcomes that are not predictable from the behavior of any individual alone. Through relational event simulations, researchers can study the effects of different network structures, parameters, and mechanisms on the emergence of social phenomena. By leveraging the rich data available on relational events and using simulation techniques, researchers can gain new insights into the mechanisms underlying emergent phenomena and develop more accurate and effective theories.

\textit{Evaluate boundary conditions of theories}
Evaluating boundary conditions, or scenarios under which a theory makes sense is critical for developing and advancing theories. Unfortunately, little is known about the boundary conditions of most social science theories. Knowing the boundary conditions of a specific theory can help guide setting up proper empirical experiments or the appropriate use of the theory in an empirical study. In our context, boundary conditions could refer to ranges of the parameter space of network effects, constraints on values of statistics, size of the network, or exogenous conditions that need to be present for a theory to be applicable. Also, knowing that a particular theory only can make valid predictions into the short future or only under fairly stable conditions informs the researcher how far into the future one can predict based on a fitted model or how much history must be taken into account in predicting future relational events in accordance with a specific theory. Simulations can be a powerful tool to explore the applicability of theories under various such conditions. 

The approach here is to take a specific theory about social interaction that one intends to use in a research project and translate that into a relational event model. By simulating network interaction patterns across a range of parameter values or initial conditions, it can be assessed beyond which (combinations of) values the model starts to generate non-sensical or unrealistic dynamics such as the emergence of strongly separated subgroups, unrealistically high density, hyperactivity of actors with certain traits, odd speeding up and slowing down of interaction rates, etc. When appropriate characteristics of the simulated relational event network no longer pass the relevant sanity checks (either by being obviously non-sensical or by being quite removed from the dynamics of empirical reference data), it becomes clear that the theory that is mimicked by the relational event simulation no longer is realistic beyond these (combinations of) parameter values or initial conditions. This, then, provides the boundary within which a researcher would like to stay when using the theory to explain a phenomenon of interest.


\textit{Incorporate timing and dynamism}. 
The proposed simulation frameworks allow researchers to incorporate dynamism when developing theories by including feedback loops that are realized through endogenous statistics, time-varying exogenous covariates when attributes of actors and dyads can vary with time, time-varying parameters that can simulate realistic change-points, and memory when the influence of past events on new events decays with time. One advantage of this is that a theorist can assess what happens when a static social theory (which almost all established social theories to date are) is reformulated in an explicitly dynamic, time-sensitive manner. Although many of our current social theories are built on inherently dynamic ideas and arguments, they are rarely formulated (and validated) as such. It is often not at all obvious how to turn established static social theories into time-sensitive theories, since we have very little knowledge of how fast interactions develop, how long it takes until routine sets in, or how long emergent properties last. Simulating several dynamic versions of current social theories can provide hints to a theorist as to how an explicitly time-sensitive and dynamic version of the theory could be developed and put to the test.

\subsection{Group formation using Optimal Distinctiveness Theory}
As an example, we present a simulation approach to represent the Optimal Distinctiveness Theory using relational event simulations. The Optimal Distinctiveness Theory of \cite{brewer_1991_SocialSelfBeinga} dictates that individuals have two fundamental and competing needs: 1) the need for inclusion, i.e., a desire to assimilate and interact with other individuals who share their social identity and 2) the need for distinctiveness from others in the actor's surroundings. Individuals prefer to be identified with social groups that are neither too inclusive nor too distinctive, but are of optimal distinctiveness.  These competing psychological mechanisms that operate at the actor level motivate the emergence of social groups that satisfy both these needs at the same time \citep{leonardelli_2010_OptimalDistinctivenessTheory}. Drawing upon optimal distinctiveness theory, we design an actor-based simulation experiment to observe the emergence of groups and their dynamics based on actor's competing needs for optimal distinctiveness. We use an actor-orientated simulation approach, keeping in mind that the actor retains agency to decide if and when to send events to other individuals in the network in accordance with the theory. Our simulation model assumes that actors make decisions about when to send an event in an attempt to satisfy their preference for interactions that are optimally distinct. Each actor $i$ has an attribute $z_i$ (that reflects their fixed identity) and a preference for the optimal value of distinctiveness associated with that actor $d^{*}_i$. This value reflects the desired proportion of actors that $i$ interacts with who do not share the same identity as $i$. 

The distinctiveness of an actor $d_i$ at any time $t$ is quantified with a proportion of distinct events, i.e. the total number of incoming and outgoing events that the actor $i$ sends and receives from other actors who have an attribute value different from $i$'s divided by the total number of events sent and received by $i$ from all other actors irrespective of the attribute:
\begin{equation}
d_i(t) = \dfrac{\sum \limits_{j \in \mathcal{A}/\{i\} } ( a_{ij}(t)+a_{ji}(t) ) \  \mathbb{I}[z_i \neq z_j ]  } {\sum \limits_{j \in \mathcal{A}/\{i\} } ( a_{ij}(t)+a_{ji}(t) )},
\end{equation}
where $a_{ij}(t)$ is the corresponding element of an $n \times n$ adjacency matrix $A(t)$ containing the total number of past events between actors at time $t$. Recall from Section \ref{sec2:dynam} that the DyNAM model requires two separate statistics for the sender and receiver choice respectively. The distinctiveness aspect of an actor's interaction preferences is incorporated into the simulation framework as a combination of a sender dissatisfaction statistic $X^s$ and a receiver choice statistic $X^r$.  

Specifically, we operationalize, a sender $i$'s dissatisfaction statistic $X^s(i,\mathcal{E}_t)$ as the absolute difference between the current distinctiveness and the optimal value of distinctiveness the actor strives for $| d_i (t) - d^*_i |$. A positive effect of the distinctiveness statistics implies that actors who are most dissatisfied with their current level of distinctiveness are most likely to be activated as the next sender in an attempt to adjust their distinctiveness level. We assume that for any actor, the direction of deviation from their optimal distinctiveness, i.e towards more or less distinctiveness does not influence the absolute level of dissatisfaction. Moreover, dissatisfied actors with a lower distinctiveness than their optimal level ($d_i (t) < d^*_i$) are likely to reach out to actors with attribute levels distinct from their own. Similarly, dissatisfied actors with a higher distinctiveness than their optimal level ($d_i (t) > d^*_i$) are likely to reach out to actors with the same attribute level as their own. If all the actor's are equally satisfied with their current distinctiveness level, an equilibrium will be maintained where all actors are equally likely to be sampled as the next sender \citep{leonardelli_2010_OptimalDistinctivenessTheory}.

For the receiver choice model of the DyNAM, a receiver choice  statistic can be defined as follows:

\begin{equation} \label{eqn:theory-ex-recv}
X^r (j | i, \mathcal{E}_{t_m}) = \begin{cases} 
    \mathbb{I}[z_i = z_j]  & \ d_i (t) > d^*_i \\
    \mathbb{I}[z_i \neq z_j]  & d_i (t) < d^*_i
\end{cases}
\end{equation}

In case $ d_i (t) > d^*_i $, i.e., when sender $i$'s interactions are more distinct than desired, the receiver statistic assigns a value of 1 for potential receivers with attribute the same as that of actor $i$. Similarly in case $ d_i (t) < d^*_i $, i.e., when actor $i$'s interactions are more assimilated than desired, the receiver statistic assigns a value of 1 to receivers with attributes distinct from $i$'s. Otherwise the statistics is set to 0. We further include an interaction of the above-defined statistic with an inertia receiver choice statistic in our model to ensure that receivers who share a greater volume of past events with a sender are more likely to be selected by the sender. 

\begin{figure}[H]
\centering
\begin{subfigure}{0.45\linewidth}
  \centering
  \includegraphics[width=\textwidth]{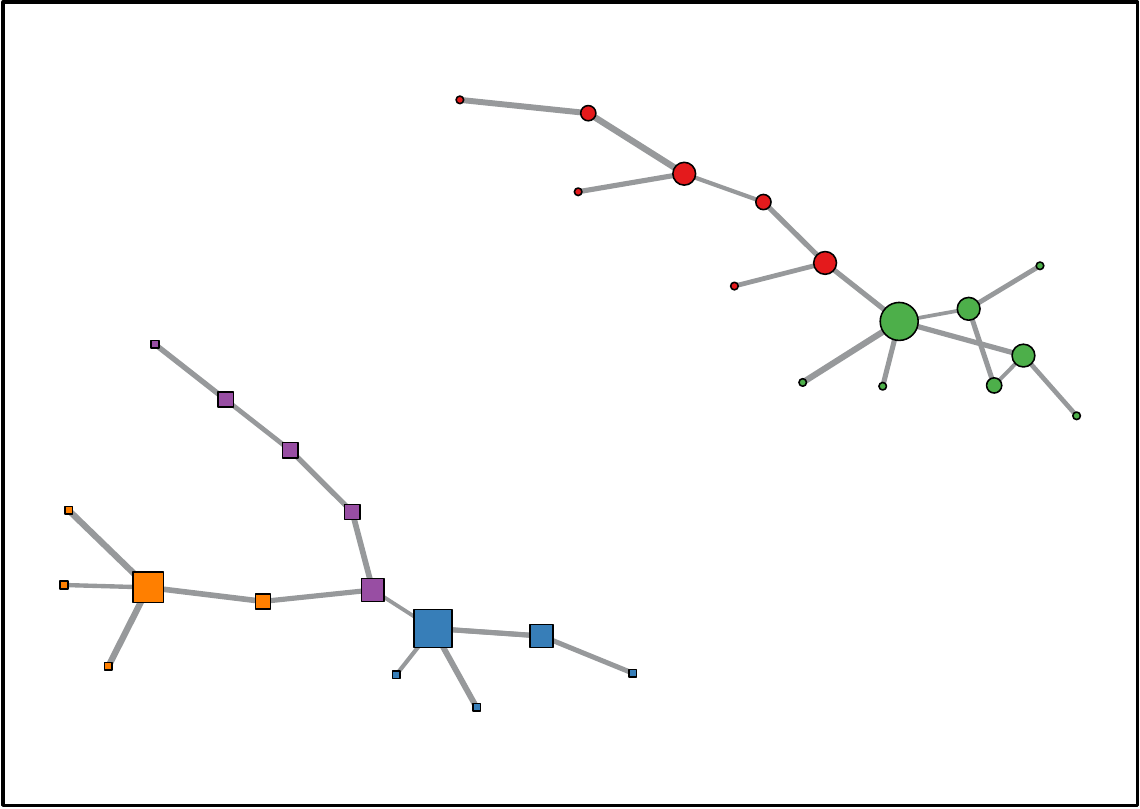}
  \caption{$d^*=0$}
  \label{fig:d0}
\end{subfigure}%
\hfill
\begin{subfigure}{0.45\linewidth}
  \centering
  \includegraphics[width=\textwidth]{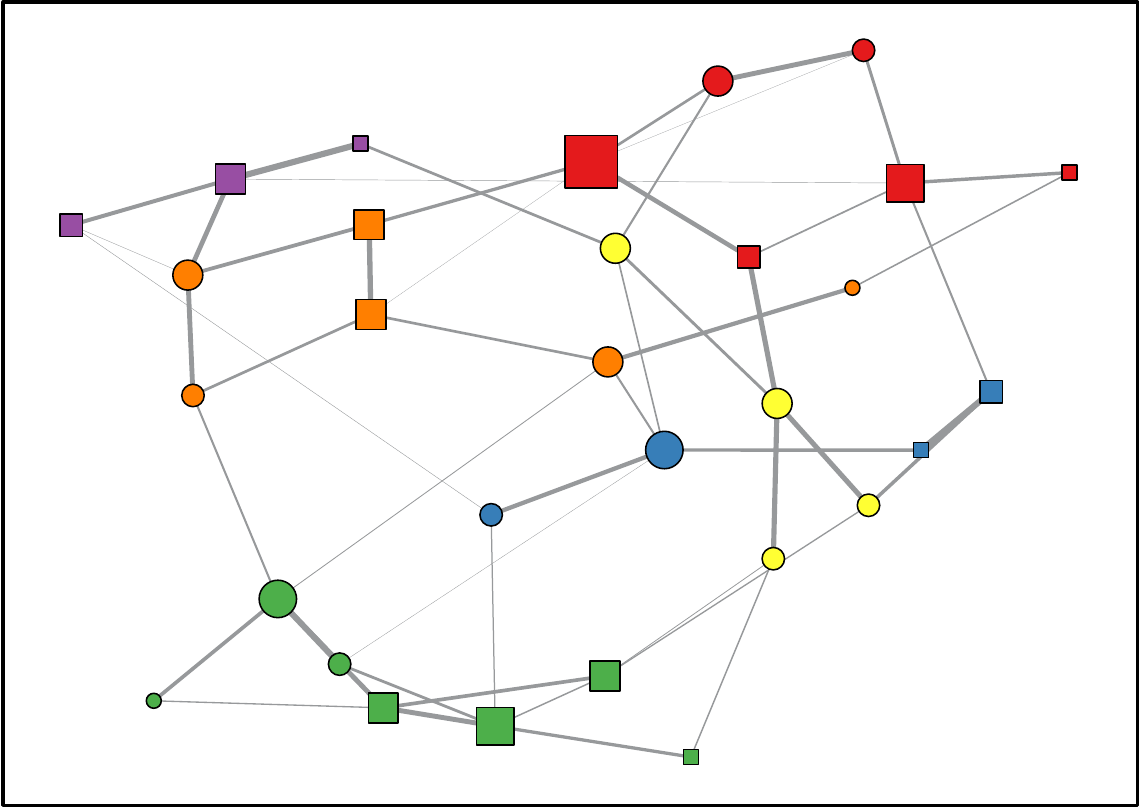}
    \caption{$d^*=0.3$}
    \label{fig:d0.3}
\end{subfigure}
\vfill
\begin{subfigure}{0.45\linewidth}
  \centering
  \includegraphics[width=\textwidth]{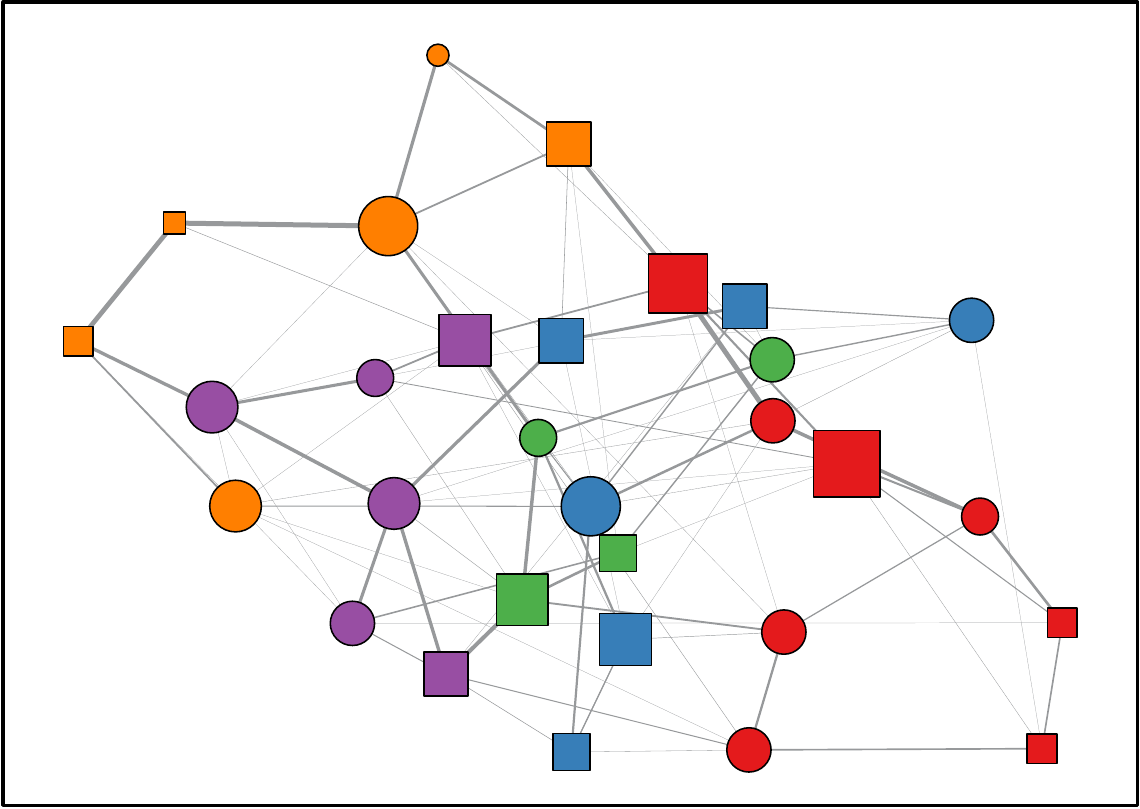}
  \caption{$d^*=0.5$}
  \label{fig:d0.5}
\end{subfigure}%
\hfill
\begin{subfigure}{0.45\linewidth}
  \centering
  \includegraphics[width=\textwidth]{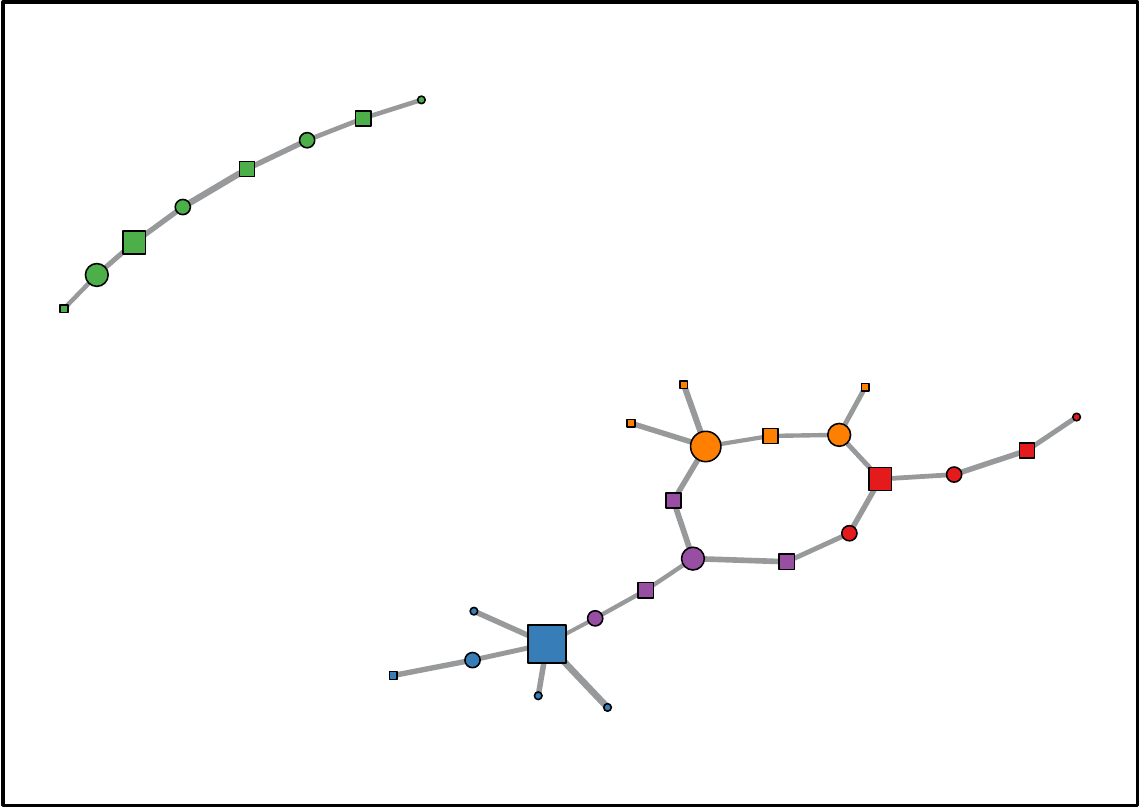}
  \caption{$d^*=1$}
  \label{fig:d1}
\end{subfigure}%
\caption{Simulated Network plots for different optimal distinctiveness $d^*$ values. Square nodes represent nodes with $z=1$ and circular nodes represent nodes with attribute value $z=2$. The colours indicate the emergent groups to which the nodes were assigned based on the community detection algorithm. The size of nodes corresponds to the degree of the actors, and the width of the edges corresponds to the total volume of events exchanged in either direction between the two nodes.}
\label{fig:theory-ex-graphs}
\end{figure}

In our simulations, we assume that all actors have the same value of optimal distinctiveness $d^*$. We simulate networks consisting of 30 actors with a binary attribute that represents their fixed identity $z=\{1,0\}$  for varying values of $d^*$. The proportion of actors with each attribute is equal. To identify emergent groups formed as a result of the simulations, we use the Louvain community detection algorithm \citep{blondel_2008_FastUnfoldingCommunities}. This method of community detection essentially partitions nodes based on a greedy modularity optimisation. The groups identified from this method are considered emergent because they arise from interactions and decisions made by actors on the actor level which lead to emergent group formation on a network level. Figure \ref{fig:theory-ex-graphs} depicts the network plots for simulated sequences with different $d^*$ values. When $d^* = 0$ as illustrated in Figure \ref{fig:theory-ex-graphs} (\subref{fig:d0}), actors in the simulation strongly prefer homophilic interactions, resulting in groups containing nodes of one attribute type. Actors with opposing attributes tend to cluster together rather than interact with actors of a different attribute.  

For $d^* = 0.3$, shown in Figure \ref{fig:theory-ex-graphs} (\subref{fig:d0.3}), a mixed grouping is observed, with some larger groups containing both attribute types and some groups being homogeneous. Actors of opposing attributes interact with each other more often, and the distance between them is shorter compared to the plot for $d^* = 0$. Figure \ref{fig:theory-ex-graphs} (\subref{fig:d0.5}) for  $d^* = 0.5$ shows a mixed grouping with equal proportion of squares and circles in the groups as well as a roughly random communication structure. The actors form a highly connected network, and actors with opposing attributes are roughly as close to each other as actors with the same attribute. In Figure \ref{fig:theory-ex-graphs} (\subref{fig:d1}) where $d^* = 1$, there is a strong preference for distinctiveness among actors in their interactions, leading to a plot with mixed groupings. However, it is worth noting that the interactions within each group are bipartite, meaning that squares are only strongly associated with circles and vice versa.

These network plots pass our sanity checks, indicating the specific way we translated the optimal distinctiveness theory into the relational event model is sound and does not produce unexpected results. We can further test the boundary conditions of the theory through simulations. In the previous simulations, the attributes were equally distributed between the actors; however, this may not often be realistic. There may be an imbalance in the proportion of actors for instance when dealing with integration of minority communities or under-represented sections of society at a university. To understand the effects of proportions of actors with different attributes on the network under the optimal distinctiveness theory, we simulated networks with 50 actors, and varied the proportion of the minority within the population. The simulations were repeated 50 times for each combination of proportion of minority actors $p \in \{0.1,0.2,0.3,0.4,0.5\}$ and for each $d^*$ values in the interval $[0,1]$.

\begin{figure}[ht]
\centering
\begin{subfigure}{0.5\linewidth}
  \centering
  \includegraphics[width=\textwidth]{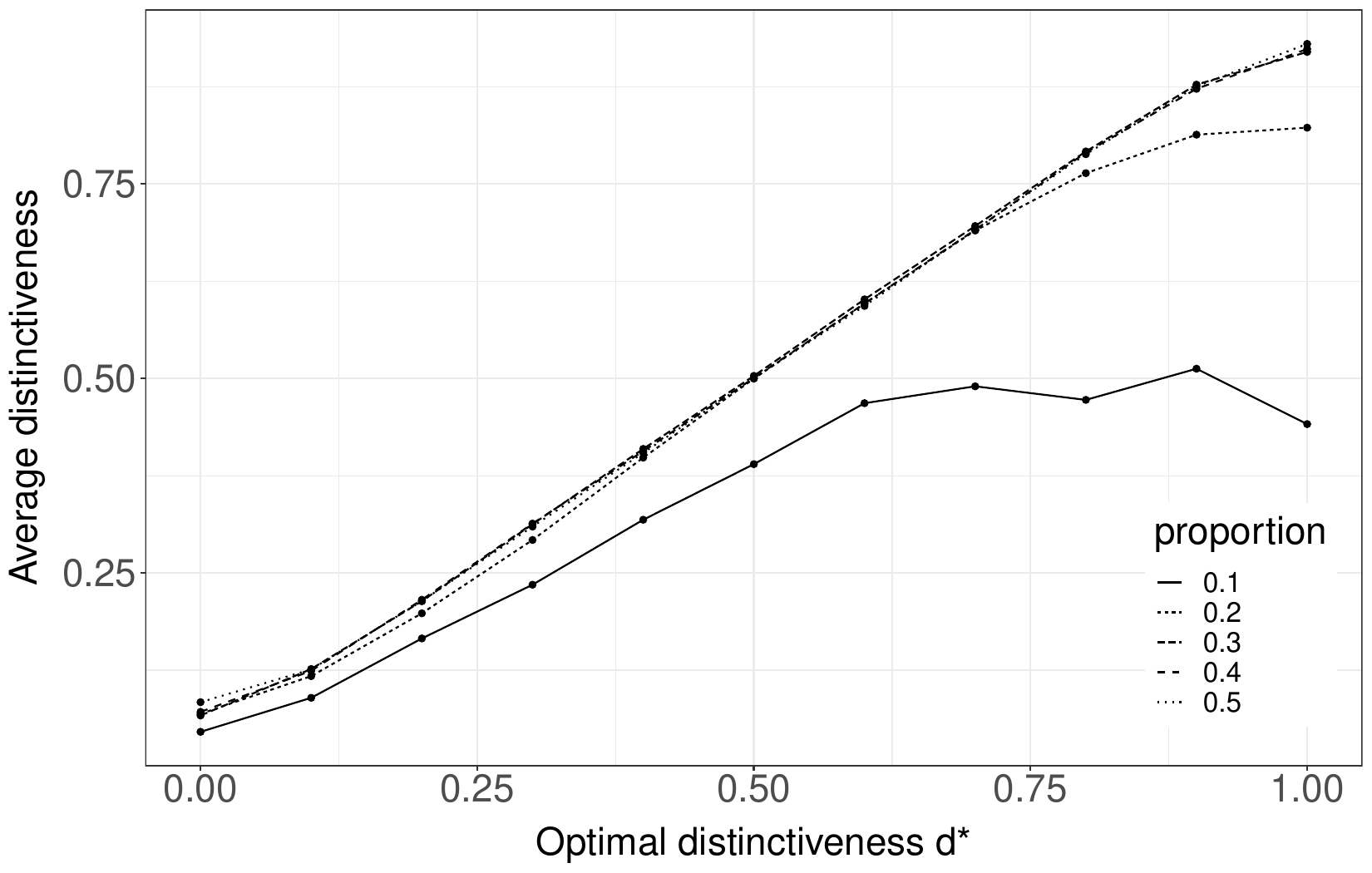}
    \caption{Average distinctiveness}
  \label{fig:avg-d}
\end{subfigure}%
\hfill
\begin{subfigure}{0.5\linewidth}
  \centering
  \includegraphics[width=\textwidth]{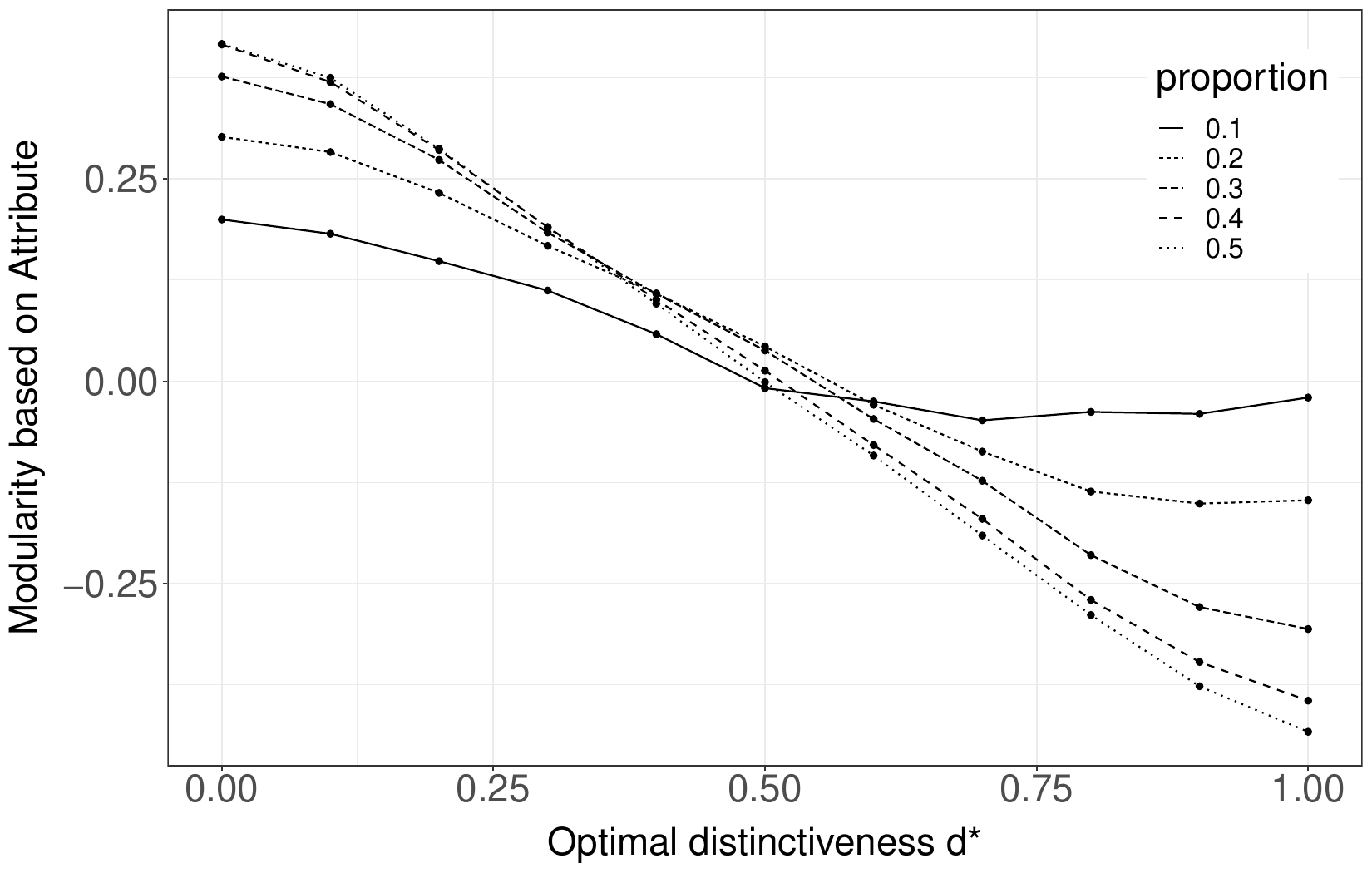}
    \caption{Average modularity based on Attributes}
  \label{fig:mod-attr}
\end{subfigure}%
\hfill
\begin{subfigure}{0.5\linewidth}
  \centering
  \includegraphics[width=\textwidth]{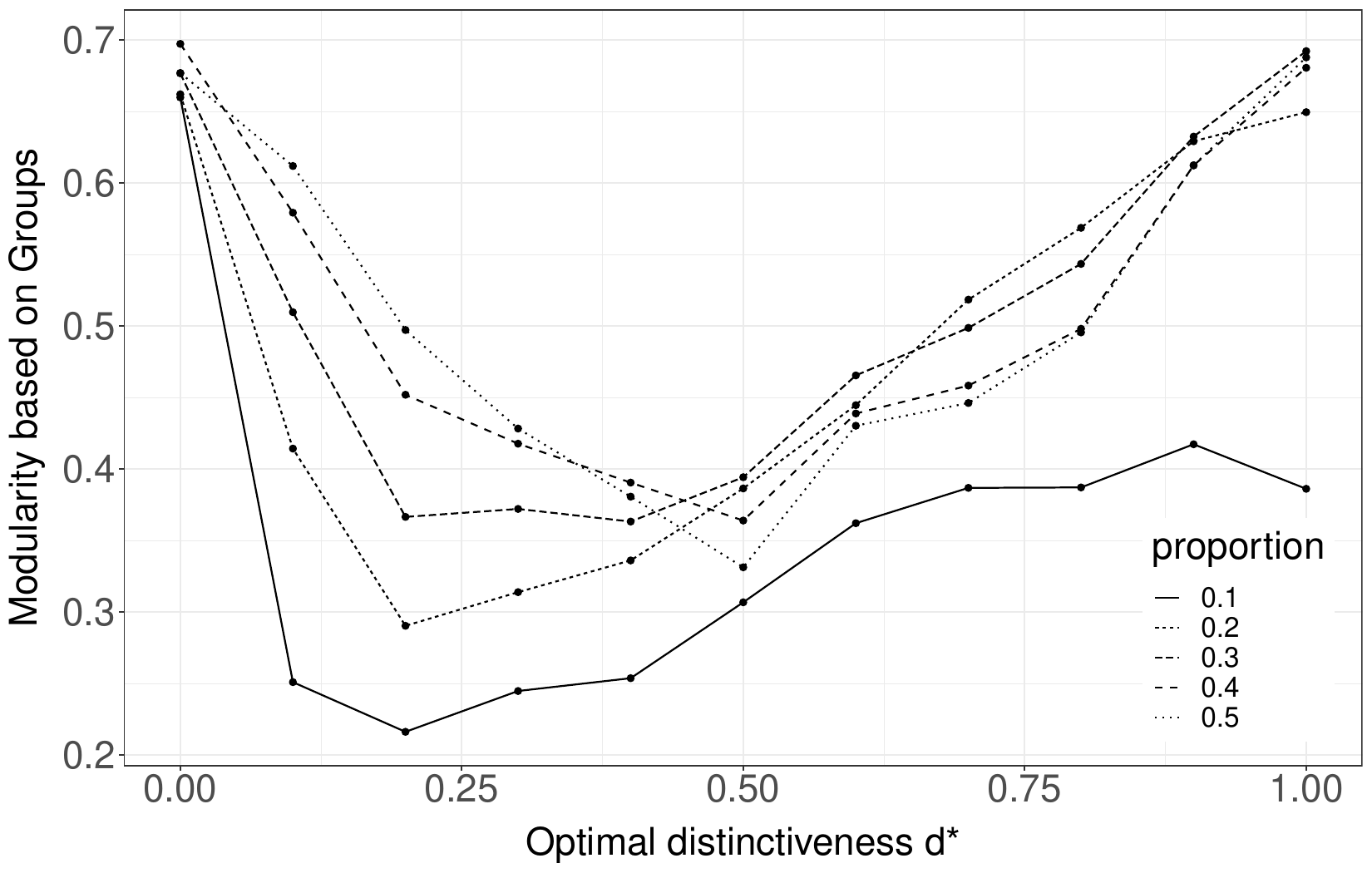}
    \caption{Average modularity based on groups}
  \label{fig:mod-comm}
\end{subfigure}
\caption{(a) Average distinctiveness of actors, (b) Average modularity computed based on the partitioning of nodes by actor attribute and, (c) Average modularity computed based on the partitioning of nodes by the detected groups at the end of 50 simulation runs for each value of optimal distinctiveness $d^* \in [0,1]$ and for each proportion of minority attribute.}
\label{fig:theory-ex-graphs-prop}
\end{figure}

Figure \ref{fig:theory-ex-graphs-prop} (\subref{fig:avg-d}) reports the average distinctiveness at the end of the simulations for the simulated event histories for the corresponding $p$ and $d*$ values. The results indicate that the average distinctiveness of the actors at the end of the simulations could converge to the optimal value when the proportions of the minority are greater than $10\%$. However, when $p=0.1$ the actors do not achieve their desired distinctiveness. When a minority is too small in size relative to the population, the actors are unsatisfied in their need for distinctiveness particularly when $d^*$ is high.

Figure \ref{fig:theory-ex-graphs-prop} (\subref{fig:mod-attr}) presents the modularity values \citep{newman_2006_ModularityCommunityStructure} computed based on the partitioning of nodes by actor attribute at the end of the simulations. The value of the modularity reflects a measure of the strength of the partitions based on actor attributes, i.e. when modularity is high actors with the same attributes tend to cluster together and have a denser communication structure. The results show that with increasing $d^*$, the modularity values decrease. Indicating that at $d^*=0$ i.e. the actors in the simulation preferred only homophilic interactions, the simulated network indeed has high modularity. Whereas at $d^*=1$, when the preference for distinct interactions is at its peak, the modularity value is at its lowest.  The decrease in modularity from $d^*=0$ to $d^*=1$ is less severe for lower proportions of minorities.

Figure \ref{fig:theory-ex-graphs-prop} (\subref{fig:mod-comm}) on the other hand, presents modularity values computed based on partitioning of nodes, based on the groups detected by the community detection algorithm. When $d^*=0.5$ the modularity is at its lowest because actors don't have a preference for distinctiveness. However on each extreme i.e $d^*=$ 1 or 0 the clustering is high because actors have a strong preference for interacting with actors with specific attributes leading to tighter group formation. The modularity values for proportion $=0.5$ is fairly symmetric around $d^*=0.5$ however for lower proportion, the symmetry is altered with the minimum shifted to lower $d^*$ values. A possible explanation of this is that when $d^*$ is higher and proportion of minority actors present in the network are lower, the clustering within a group is not strong due to the majority actors fulfilling their desired for high distinctiveness from minority actors outside the group. The minimum at lower $d^*$ and lower $p$ values occurs when the groups are formed around minority actors, thus the majority actors are able to fulfill their distinctiveness criteria within their groups.

Our purpose was to illustrate through a simple example \textit{how} an interested researcher could articulate theories using the REM simulation framework. The results indicate how individual preferences of actors, in this case their desired distinctiveness, led to emergent group formation on the network level. We further tested a simple boundary for the theory by varying the proportion of minorities in the network. We see that dissatisfaction across actors is greater when the size of the minority group is skewed toward a very small minority presence compared to societies with larger minority groups, where actors are able to achieve their desired optimal distinctiveness.  Future work remains to be done to investigate the impact of including more than two attributes or when varying the optimal distinctiveness in the population where certain actors have a higher desire for distinctiveness than others. Further we assumed that the direction of dissatisfaction does not influence who is likely to be the next sender. We did not distinguish between actors dissatisfied with their distinctiveness and actors dissatisfied with their assimilation as long as the absolute value of dissatisfaction was equal. In reality, it is possible that these two cases may have a different impact on the probability of the next sender. Future work  is needed to evaluate this. In addition, we also assumed that all actors had free choice and no restrictions in their communications, however, in reality spatial or cultural restrictions could influence the interactions of actors in different ways and needs to be taken into account. 

\section{Planning network interventions via simulations}\label{sec-intv}
Network interventions describe ``the process of using social network data to accelerate behavior change or improve organizational performance'' \citep{valente_2012_NetworkInterventions}. Interventions involve the use of social network data to bring about a behaviour change or to influence social dynamics in a network towards desirable outcomes. Previous work in network interventions focuses on adoption of behaviour and transmission of the behaviour across neighbours of nodes in a network often using diffusion models \citep{valente_2005_NetworkModelsMethods,valente_2012_NetworkInterventions,valente_2017_PuttingNetworkNetwork,badham_2021_NetworkStructureInfluence}. These models assume that behaviour is propagated in a network and once a node adopts a behaviour, it can influence it's neighbour to adopt that behaviour. The edges between actors in such a network can represent dissemination of information, opinions, ideas, goods or even diseases. The variables of interest in diffusion models are often the proportion of nodes that have adopted a desirable behaviour or have acquired certain knowledge. However, in relational event models, the variable of interest is the rate of interaction that describes who interacts with whom and when. REM or DyNAM can be utilized to model interventions in evolving interaction networks where the intervention aims to influence the interaction dynamics itself.

The relational event simulation framework can be availed to predict or better understand the potential impact of the intervention outcomes over time. This can be of great practical use, for instance by allowing managers to try out several interventions ``in-silico'' before deciding which one(s) to try out  ``in-practice''; the effects of competing interventions can be assessed without spending much money or other resources and without overhauling an organization based on a mere managerial hunch. Rather, a manager can get a fairly good idea of which intervention is likely to work, how it should be implemented, and what the pitfalls are that need to be monitored when the intervention is implemented in real life. 

Ideally, the simulation is initialised from an estimated model based on an empirical observed network. This allows a realistic simulation of the intervention based on empirically derived conditions and effect estimates. \cite{adams_2016_HowInitialPrevalence} explored how the same intervention can have different results based on the initial conditions of the network. This is also applicable in the REM context. For example, consider an intervention that is conducted for a while and aims to promote inter-departmental collaboration in an organizational network. If the network is characterized by high tendencies towards inertia, the inter-department interactions that were originally triggered by the intervention (that only ran for a limited time) may become sustained into the long term due to the network's tendency for inertia. On the other hand, if inertia is only low (or negative) in this network, it is possible that, once the intervention ends, the network reverts back to its original state. Having an estimate of the effects operational in the empirical network before-hand would be highly beneficial in planning interventions. Simulations could then be carried out to evaluate the outcomes of planned interventions that explore the implications of a manipulated set of effects or initial conditions.

\subsection{Types of network interventions}
In order to effectively simulate realistic interventions using the simulation framework, it is important to understand the types of interventions that can be simulated in the relational event context and how the mathematical modification to the model translates to implementation in practise. To be able to effectively translate an intervention into the REM or DyNAM model, it is important to define which aspects of the model the intervention modifies. To facilitate this, we distinguish between three types of REM interventions: i) Actor Attributes, ii) Network Effects, and iii) Composition - based on the locus of change in the model.

\begin{itemize}
    \item Attributes: Interventions may involve modifications to attributes of actors or relationship between actors (dyadic attributes) to answer questions such as: can changing the layout of desks lead to a change in communication patterns of students in a classroom? What is the effect of changing the gender or hierarchy distribution on the communication in an organisation? The analytical translation of such interventions are straightforward because they involve simply modifying the actor or dyadic attributes while keeping other aspects of the model the same as pre-intervention.
    
    \item Network Effects: Interventions of this type focus on modifying network effects of the model itself by either changing the magnitude of network effects (by adjusting the corresponding parameters) or by removing (or adding) network effects that may (or may not) be operative in the social network previously. For example, consider an intervention that is carried out by organising social events designed to increase mixed-gender contact opportunities among university students. The analytical translation of this intervention could be a temporary increase in the magnitude of the gender heterophily effect in the duration of the organised events. Simulating events beyond the intervention duration could reveal the long-term effects of the designed intervention on ties between the students. Another example of this type of intervention is reduction in popularity of a smoker as students in a university are made aware of the harmful effects of smoking. The translation of this effect could be a decrease in magnitude of the effect for the smoker popularity statistic (interaction of the smoker attribute and the in-degree of the receiver).
    
    \item Structure: Interventions that change the structure of a social network involve changes to size of the network by addition or removal of actors, the riskset by restricting which pairs of actors may interact, and change in the composition and grouping  of actors. Examples of structural changes include addition of employees when companies merge or acquire one another, removal of a key leader in an organisation, and reassignment of teams and managers when new projects are started. The analytical translation of these interventions are straightforward and often involve making alterations to the riskset. In case of re-assignment of teams for instance, dyads with employees from different teams or locations may no longer be considered in the riskset. 
    
\end{itemize}

The three categories above are not disjoint, as interventions in practise can involve multiple categories. For example, consider an intervention that involves the addition of sport activities in a university. Students and staff from different departments would now have an opportunity to interact with each other during these events. The analytical translation of such an intervention involves all the three categorizations defined above. Actor attributes that represent whether or not individual actors participate in sports and the type of sport if they do, are added to the model. A new network effect can be added for homophily interactions that may take place between actors attending the same sports sessions. Additionally, the structure of the riskset may also be altered to include dyads from different departments that participate in the same sports that may not previously be considered at risk of interaction. 

\begin{figure}[t]
\centering
\begin{tikzpicture}[scale=0.75]
\draw[thick, -Triangle] (0,0) -- (\ImageWidth,0) node[font=\scriptsize,below left=3pt and -8pt]{time};

\foreach \x in {0,1,...,10}
\draw (\x cm,3pt) -- (\x cm,-3pt);

\foreach \x/\descr in {0/t_0,
3/t_1,6/t_2,8/t_3}
\node[font=\scriptsize, text height=1.75ex,
text depth=.5ex] at (\x,-.3) {$\descr$};

\draw [thick ,decorate,decoration={brace,amplitude=5pt}] (0,0.5)  -- +(3,0) 
       node [black,midway,align=center,above=4pt, font=\scriptsize] {Pre-Intervention\\$\boldsymbol{\beta}(t)_{het} = \boldsymbol{\beta}(t)_1$};
\draw [thick,decorate,decoration={brace,amplitude=5pt}] (6,-.7) -- +(-3,0)
       node [black,midway,align=center,font=\scriptsize, below=4pt] {Intervention I\\$\boldsymbol{\beta}(t)_{het} = \boldsymbol{\beta}(t)_2$};
\draw [thick ,decorate,decoration={brace,amplitude=5pt}] (6,0.5)  -- +(2,0) 
       node [black,midway,align=center,above=4pt, font=\scriptsize] {Post-Intervention I\\$\boldsymbol{\beta}(t)_{het} = \boldsymbol{\beta}(t)_1$};
\end{tikzpicture}

\begin{tikzpicture}[scale=0.75]
\draw[thick, -Triangle] (0,0) -- (\ImageWidth,0) node[font=\scriptsize,below left=3pt and -8pt]{time};

\foreach \x in {0,1,...,10}
\draw (\x cm,3pt) -- (\x cm,-3pt);

\foreach \x/\descr in {0/t_0,
3/t_1,7/t'_2,10/t'_3}
\node[font=\scriptsize, text height=1.75ex,
text depth=.5ex] at (\x,-.3) {$\descr$};

\draw [thick ,decorate,decoration={brace,amplitude=5pt}] (0,0.5)  -- +(3,0) 
       node [black,midway,align=center,above=4pt, font=\scriptsize] {Pre-Intervention\\$\boldsymbol{\beta}(t)_{het} = \boldsymbol{\beta}(t)_1$};
\draw [thick,decorate,decoration={brace,amplitude=5pt}] (7,-.7) -- +(-4,0)
       node [black,midway,align=center,font=\scriptsize, below=4pt] {Intervention II\\$\boldsymbol{\beta}(t)_{het} = \boldsymbol{\beta}(t)_3$};
\draw [thick ,decorate,decoration={brace,amplitude=5pt}] (7,0.5)  -- +(3,0) 
       node [black,midway,align=center,above=4pt, font=\scriptsize] {Post-Intervention II\\$\boldsymbol{\beta}(t)_{het} = \boldsymbol{\beta}(t)_1$};

\end{tikzpicture}
\caption{Diagrammatic representation of two interventions}
\label{fig:intv1}
\end{figure}
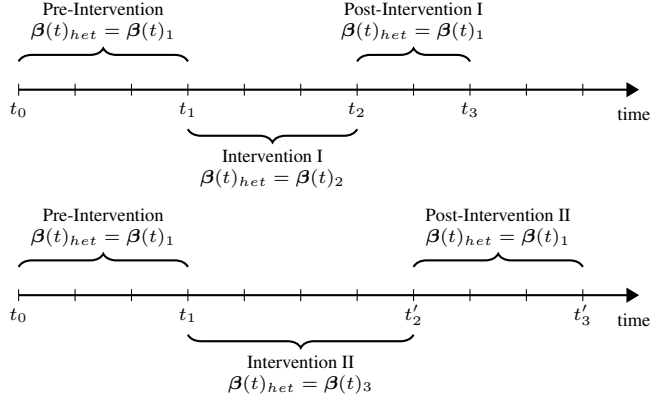
\subsection{Evaluating the Persistence of Intervention Outcomes}
In this example, our aim is to explore the longevity of outcomes of different interventions through simulations. With simulations, it is possible to assess how a network would react to competing interventions and assists in improving intended network interventions towards a desirable outcome. The intervention in this example is intended to increase the inter-department collaboration between employees of two departments in an organization. The number of inter-departmental emails is used as a proxy for the collaboration across departments. This intervention example is a simplified version of an actual intervention study that we conducted for the R\&D department of a large consumer goods company in Western Europe. To increase the inter-department communication, the organisation intended to intervene by providing the employees with incentives to increase their inter-department communication (e.g., by organising new multi-department projects and organising joint meetings). This intervention is modelled in our example by temporarily increasing the departmental heterophily effect $\boldsymbol{\beta}(t)_{het}$ in the model. Our interest is in evaluating what the effect is of the intervention after it ends and how long the increased heterophily effect remains in the network after the intervention has ended and the network has returned to its ``normal" dynamic. Our assumption is that the intervention by itself will not structurally affect employee interaction norms and preferences. Therefore, after the intervention, we assume that the ``pre-intervention" model is again applicable to the interaction among the employees (but the interaction histories will have been altered by the interventions).

 Before testing the interventions using simulations, we first estimate the coefficients of a model based on interactions before the intervention, for a period spanning 243 days using a standard REM. The REM model contains various endogenous effects (represented by $\boldsymbol{\beta}(t)_{endo}$) such as inertia, reciprocity, participation shifts and degree effects. Exogenous effects (represented by $\boldsymbol{\beta}(t)_{exo}$) for the gender, seniority and sub-domain are also included. An additional effect for inter-departmental heterophily ($\boldsymbol{\beta}(t)_{het}$ is also included and is the locus for the intervention. 

The model fits the data reasonably well and we assume that it captures the way in which the company's employees tend to interact well. In the following, we will compare nine versions of the intervention to increase the inter-departmental homophily (by increasing $\boldsymbol{\beta}(t)_{het}$). The interventions differ in the length of time they run and in the strength of the intervention and we aim to evaluate the persistence of the intervention effects after the intervention has been carried out. Figure \ref{fig:intv1} graphically shows two examples. Intervention I lasts from $t_1$ to $t_2$ and for this period we increase $\boldsymbol{\beta}(t)_{het}$ (the parameter for the heterophily effect) to $\boldsymbol{\beta}(t)_2$. Alternatively, intervention II runs longer than intervention I (from $t_1$ to $t_2'$) and has a different heterophily parameter $\boldsymbol{\beta}(t)_{het}=\boldsymbol{\beta}(t)_3$. 

\begin{figure}[t]
    \centering
      \includegraphics[width=0.75\textwidth]{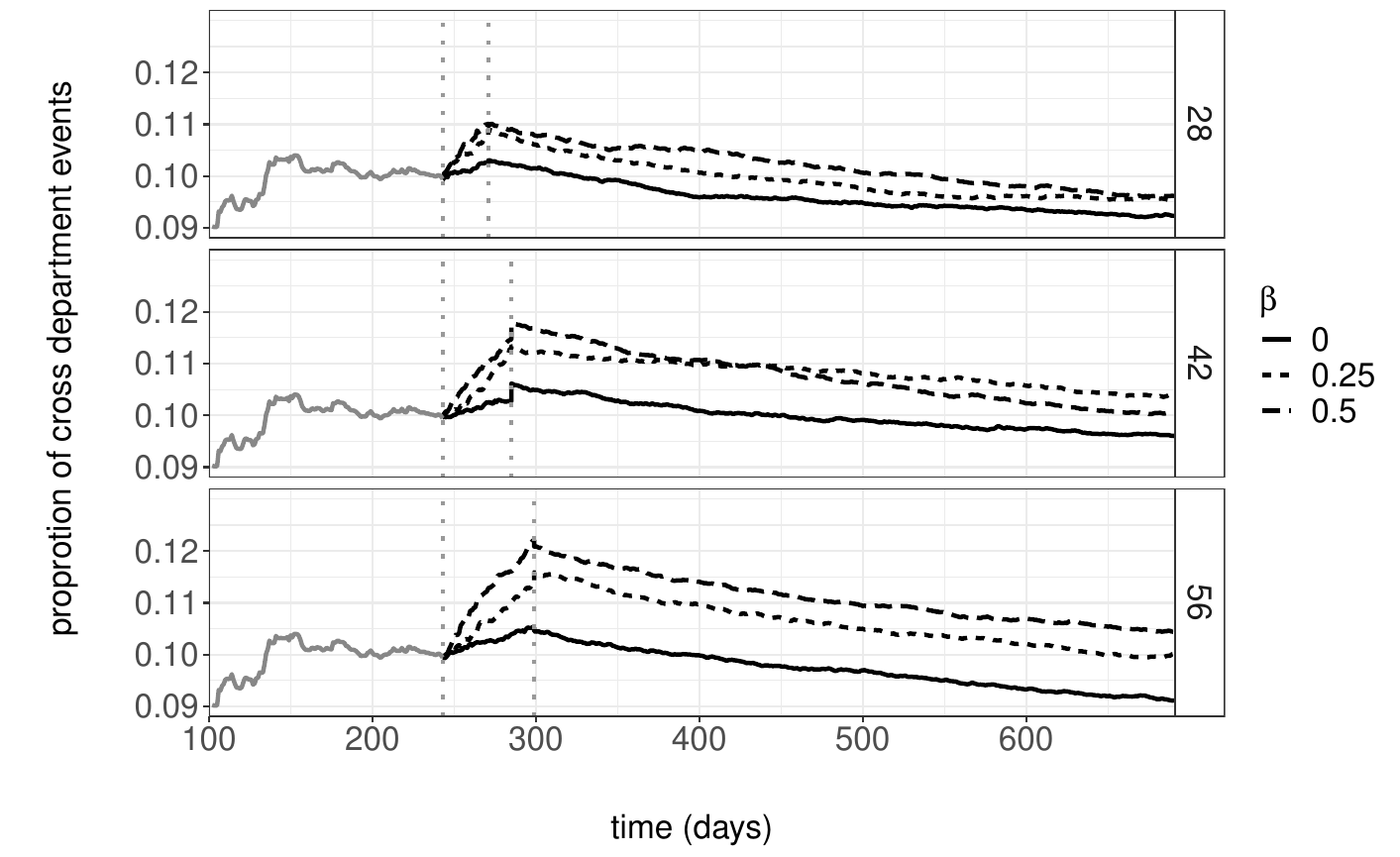}
    \caption{Median proportion of inter-departmental events simulated after an artificial intervention. The vertical dotted lines indicate the the intervention period. Top panel shows the simulations for an intervention lasting 4 weeks i.e 28 days in duration. Middle panel with duration 6 weeks i.e 42 days and Bottom panel with 8 weeks i.e 56 days.}
    \label{fig:intervention-params}
\end{figure}

Using the pre-intervention sequence as the initialization, we simulate the relational events in the network for the length of time the intervention runs and following the intervention, we further simulate the relational sequence for 50 weeks after each intervention. We consider three durations of intervention $\{28,42,56\}$ days and three inter-departmental heterophily effect sizes $\{0,0.25,0.5\}$. Therefore, we generate relational event sequences under $3 \times 3 = 9$ different interventions. Since the simulations have a stochastic component, we generate 50 relational event sequences for each intervention and compute the proportion of interdepartmental events in each. For each intervention, we report the median proportion in the 50 simulated sequences in Figure \ref{fig:intervention-params}. The proportion of inter-department events increased most for interventions held for the longest period and for the strongest effect size. The effects of the interventions on the inter-departmental communication lasted 38 days for the case of $\boldsymbol{\beta}(t)_{het}=0$, 91 days for the case of $\boldsymbol{\beta}(t)_{het}=0.25$ and 208 days when $\boldsymbol{\beta}(t)_{het}=0.5$ when the duration of the intervention period was 28 days. Similarly, the effects of the intervention last 61 days for $\boldsymbol{\beta}(t)_{het}=0$, 247 days for $\boldsymbol{\beta}(t)_{het}=0.25$, and 283 days for $\boldsymbol{\beta}(t)_{het}=0.5$ when intervention duration was 42 days. Further, the impact of the intervention lasted 72 days for $\boldsymbol{\beta}(t)_{het}=0$, 307 days for $\boldsymbol{\beta}(t)_{het}=0.25$, and persist beyond our simulation period when $\boldsymbol{\beta}(t)_{het}=0.5$ when intervention duration was 56 days. We conclude that relational event simulations can provide a useful tool in planning and designing network interventions as well as to evaluate the persistence of interventions outcomes over time.

\subsection{Evaluating Intervention Targeting Strategies}
Our second example concerns the situation where one would like to know whom to aim a specific intervention at. For example, in organisational networks, it may be prohibitively costly and complicated to intervene directly in everyone's work. Rather, it may often be effective to target an intervention at a limited set of employees only, acknowledging that their changed behaviour will subsequently affect that of their coworkers, causing the intervention effect to spread across the network naturally. Of course, the effectiveness of this will usually depend on which set of employees are targeted by the intervention. This is what we will show an example of here. Again, the intervention is intended to increase inter-departmental collaboration.

\begin{table}[t]
\def\arraystretch{1.2}
\begin{tabular}{p{7cm} p{9cm}}
\hline
\hline
\textbf{Targets} & \textbf{Description} \\
\hline
1. Random &   Each actor has equal chance of selection. \\
2. Highest Centrality & Actors with the highest betweenness centrality are selected. \\ 
3. Lowest \ Centrality & Actors with the lowest betweenness centrality are selected. \\ 
4. Highest Inter-department Out-degree & Actors with the highest number of inter-departmental out-going events are selected. \\ 
5. Lowest Inter-department Out-degree & Actors with the lowest number of inter-departmental out-going events are selected. \\ 
6. Senior & Actors designated as `Senior' in the dataset (based on their tenure at the organization) are selected.\\
7. Junior & Actors designated as `Junior' in the dataset (based on their tenure at the organization) are selected. \\
8. Male & Actors designated as `Male' in the dataset are selected.\\
9. Female & Actors designated as `Female' in the dataset are selected. \\
10. No Intervention & No intervention was carried out. \\
 \hline
\end{tabular}
\caption{Overview of measures used to select intervention targets}
\label{tab:intv-strategies}
\end{table}

Keeping the duration of the intervention constant, we consider different targets of the intervention. Table \ref{tab:intv-strategies} provides an overview of the nine different sets of targets that we consider. The ``random targets" intervention is used as a baseline to compare the other, more specifically targeted, interventions against. Interventions 2-5 use the endogenous activity of actors as a criteria for target selection. Interventions 6-9 are based on exogenous attributes of the actors. We also include a scenario when no intervention is carried out.

 The interventions in our example are carried out by modifying the department-heterophily parameter $\boldsymbol{\beta}(t)_{het}$ only for the selected targets. Specifically, the departmental heterophily parameter during the intervention remains the same as its value pre-intervention ($\boldsymbol{\beta}(t)_{het}^{pre} = -0.58$) for dyads where the sender does not belongs to the target group of actors $\mathcal{Q}$. However, the heterophily parameter is increased to a higher value ($\boldsymbol{\beta}(t)_{het}^{intv} = 0.5$) for dyads where the sender belongs to the target group of actors. Thus, the heterophily effect is split into two groups with two different parameter values.

\begin{equation*}
\centering
\boldsymbol{\beta}(t)_{het}(i,j) = \begin{cases} 
    \boldsymbol{\beta}(t)_{het}^{pre}  & \ i \notin \mathcal{Q}^{K} \\
    \boldsymbol{\beta}(t)_{het}^{intv} & \ i \in \mathcal{Q}^{K}
\end{cases}
\end{equation*}

\begin{figure}[th]
\centering
\begin{subfigure}{0.7\linewidth}
    \includegraphics[width=\textwidth]{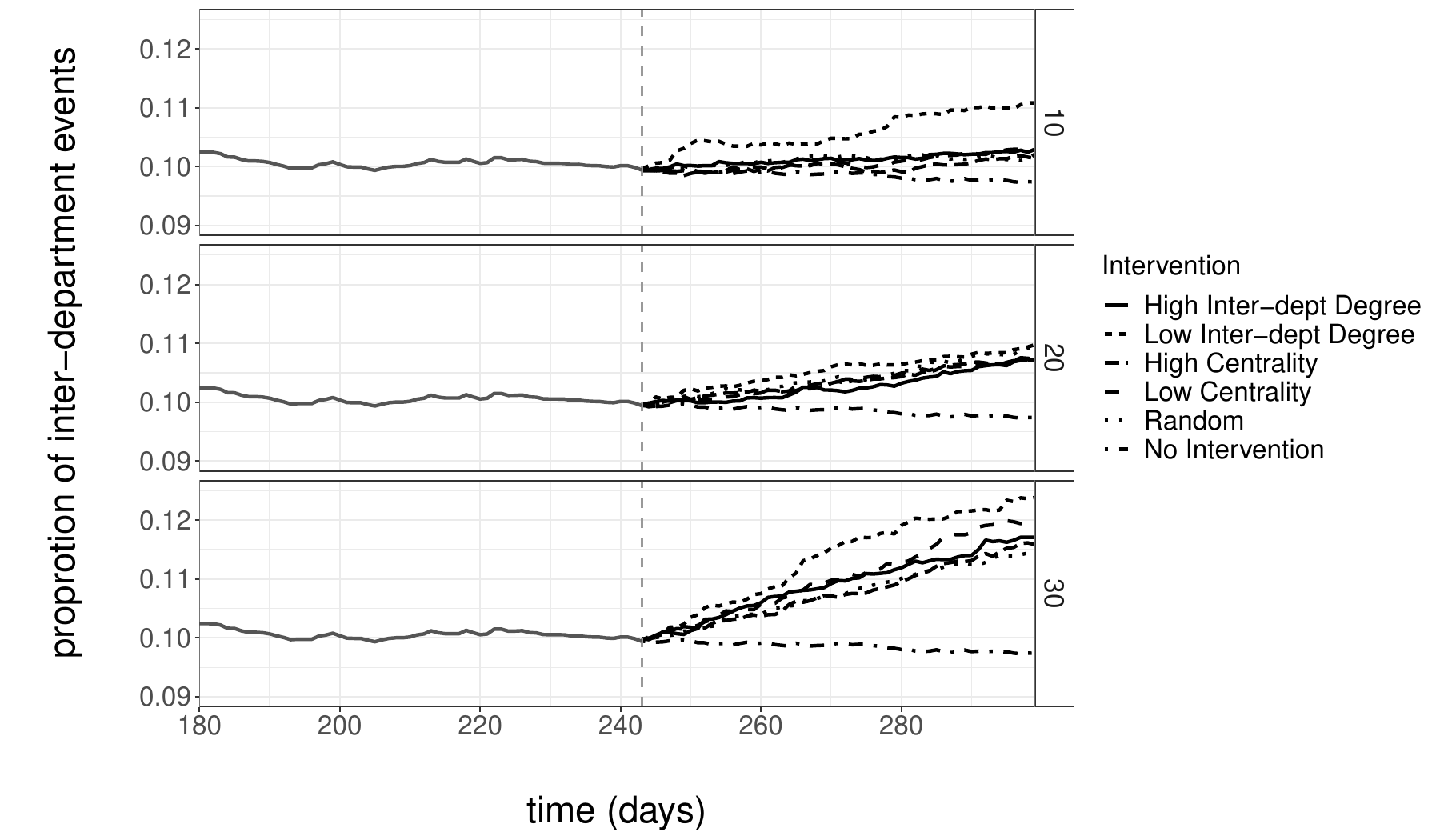}
   \caption{Intervention targets based on endogenous attributes}
   \label{fig:endo}
\end{subfigure}%
 \vfill
\begin{subfigure}{0.7\linewidth}
    \includegraphics[width=\textwidth]{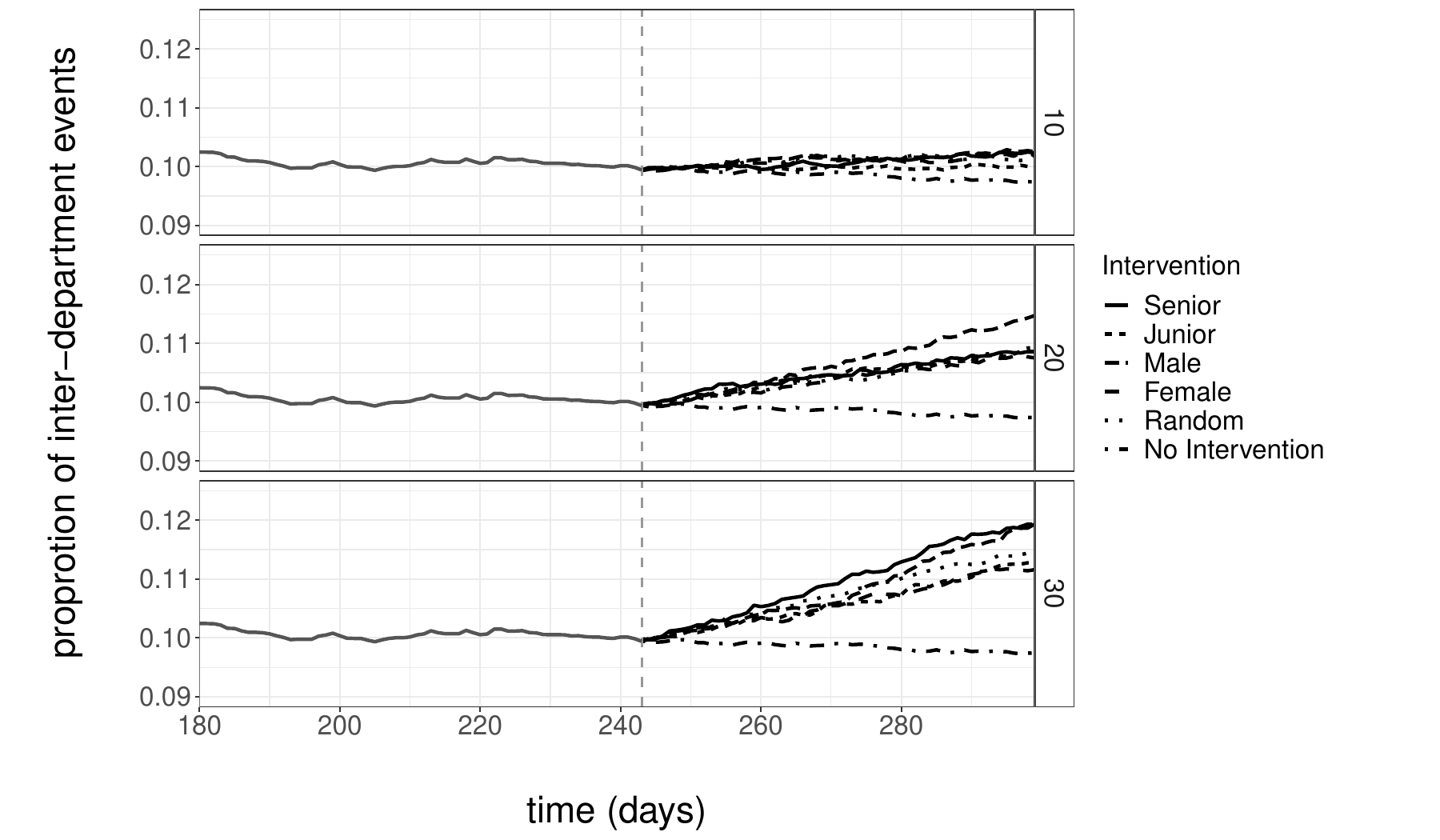}
  \caption{Intervention targets based on exogenous attributes}
  \label{fig:exo}
\end{subfigure}
\caption{Median proportion of inter-departmental events simulated for all the intervention strategies. (a) Intervention targets are selected on endogenous attributes, and (b) targets are selected on exogenous attribute. The panel in each sub-figure shows the results from top to bottom, K = \{10\% ,20\% and, 30\%\} }
\label{fig:intervention-strats}
\end{figure}

 In addition to comparing the different strategies for target selection, we also vary the fraction of actors ($K$) that are selected for each strategy (from the set of 60 employees). Figure \ref{fig:intervention-strats}(\subref{fig:endo}) shows the reported median proportion of inter-department events simulated for the endogenous selection strategies, random selection, and no intervention for K = 10\%, 20\%, and 30\%. For $K=10\%$ and $20\%$ actors selected based on having the lowest inter-department degree had the greatest impact on the proportion of inter-departmental communication while, selecting targets based on the remaining strategies impacted the proportion of inter-departmental somewhat equally as selecting random targets. When the $K = 30 \%$, the low inter-departmental degree strategy is again leading followed by the high centrality. The remaining targeting strategies impacted the proportion of inter-departmental events somewhat equally. Figure \ref{fig:intervention-strats}(\subref{fig:endo}) depicts the comparison among the exogenous attribute based strategies. Targeting seniors or females lead to the greatest impact for $K=30\%$ whereas the strategies were roughly tied when fewer fraction of actors were selected i.e $K=10\%$. In all cases targeting junior employees had least effect.

 The simulation results indicate that targeting a smaller set of actors rather than the entire network indeed can lead to desirable results in that network. However, the selection of the intervention targets can impact the results considerably. In our example, we found that targeting juniors does not produce compelling results rather seniors in the organization should be targeted for promoting inter-departmental communication. Further targeting actors who have a lower inter-departmental communication rate can prove to be more impactful than actors with higher inter-department communication or even higher centrality. Hence, when designing an intervention in practise, simulations can be an effective tool to assess which individuals to target in order to achieve more efficient results.

\section{Discussion} \label{sec6}

Relational event simulation techniques are indispensable tools for social network researchers for numerous purposes such as making predictions, building theories, testing model assumptions, evaluating hypotheses on network dynamics, or checking statistical power. To apply these techniques in a relatively easy manner, the paper presented flexible simulation frameworks under dyadic and actor-oriented relational event models which are readily available using the R package \texttt{remulate}. 
The usefuless of simulation techniques using \texttt{remulate} was shown for different types of problems in social network research. First, the package was used for assessing the goodness of fit of relational event models to study violent interactions between criminal gangs. Simulations indicated that while the model fits well in general, it may underestimate interevent time between escalations in gang conflicts, suggesting a need for model refinement to better capture these dynamics. Second, the simulation framework was used for building, evaluating, and extending social network theories, one of the most important goals of social research. This was demonstrated through the application of the Optimal Distinctiveness Theory, illustrating how individual preferences for inclusion and distinctiveness drive group formation in networks. Third, the simulation framework was used for planning network interventions in real-life, identify intervention points, and inform policy decisions. For example, simulations helped design interventions to increase inter-department communication in an organization, evaluating the persistence and impact of these intervention outcomes over time.

Although the flexible frameworks and software provide a practical means for social network researchers to simulate realistic dynamic networks for different purposes, the application of the methodology remains challenging. First, the task of specifying appropriate simulation models is not trivial and time-consuming. For this reason the use of existing theories (even on a general level), contextual understanding, and extensive exploration is essential to avoid model mis-specification. Simulations from a mis-specified model may result in unrealistic networks such as networks with unrealistically high (or low) connectivity. Network simulation techniques are indispensable to circumvent this problem. Second, the simulations we outlined in our applications only provide control over the local mechanisms of interaction. These local 'rules' give rise to global network level properties indirectly. However, our simulations do not give direct control over global network level properties. For example, if a network of a specific density or degree distribution is desired, then the methods proposed in this paper may not be helpful as the density or degree distribution of a relational event network only partially depends on the effects used to simulate relational event sequences. Despite these challenges, statistical simulation techniques belong still to the best and most flexible methods of a social network researchers’ toolkit to study social interaction dynamics in complex real-life applications. 

Worthwhile directions of future research on temporal social network simulators include relational event models that take into account more complex heterogeneity structures in network data, for instance through crossed effects models or latent variable models, more complex shapes of memory decay in endogenous statistics, and further refinement of time-sensitive goodness of fit indices. The simulation objectives and examples discussed in this article along with the easy to use open source R package \texttt{remulate} can provide the reader with a tool that opens up new avenues of research and shortens the design time for simulating time-stamped relational event data.

\section*{Acknowledgement}
The first author was supported by a NWO Grant (452-17-006). The second author was partially supported by a NWO Grant (452-17-006) and an ERC Starting Grant (758791). The third author was supported by an ERC Starting Grant (758791).

\section*{Conflict of Interest}
On behalf of all authors, the corresponding author states that there is no conflict of interest.

\section*{Data Availability Statement}   
Data utilized in Section 3.2, 5.2 and 5.3 are not publicly available. The simulation scripts are available upon request.

\newpage
\bibliography{references}

\begin{appendices}
    \section{R Code for simulations using \texttt{remulate} R package}
    \label{appendix:A}
\begin{lstlisting}[language=R]
library(devtools)
install_github("TilburgNetworkGroup/remulate")

#load the package
library(remulate)

# define number of actors for the simulations
N <- 10
# create a covariates data.frame that contains covariates for each actor in the network
cov <- data.frame(id=1:N,time=rep(0,N),gender=sample(c(0,1),N,replace=T,prob=c(0.4,0.6)),age=sample(20:30,N,replace=T))

#formula to specify statistics for tie-oriented model
effects <- ~ remulate::baseline(-3)+
    remulate::inertia(0.10,scaling="std") + 
    remulate::reciprocity(-0.04)+
    remulate::same(0.02,variable="gender", attributes = cov)+
    remulate::send(0.01, variable="age", attributes = cov)+
    remulate::interact(param=.01, indices = c(2,5)) #interaction of inertia (2) and send age (5)

#generate 100 events using tie oriented model or until time reaches a value of 100
dat <- remulate::remulateTie(effects = effects, actors = 1:N, events = 100, time = 100)

#formulae to specify statistics for actor-oriented model
rate_effects <- ~ remulate::baseline(-3) + 
    remulate::outdegreeSender(0.1)+
    remulate::send(0.01,variable="age",attributes=cov)

choice_effects <- ~remulate::inertia(0.1) +
    remulate::reciprocity(-0.3)+
    remulate::same(0.02,variable="gender",attributes = cov)

#generate using actor oriented model
dat <- remulate::remulateActor(rate_effects, choice_effects, actors = 1:N, events = 100, time = 100)

\end{lstlisting}

    \section{R code of simulations for Section \ref{sec:ex-gof}}
    \label{appendix:B}

\begin{lstlisting}[language=R]
library(remulate)
library(remstats)
library(remstimate)

reh <- remify::remify(edgelist = edgelist, model = "tie")

statistics <- ~ remstats::send(variable = "IsPubHouse", attr_actors = actor_cov)+
    remstats::send(variable = "BonPow", attr_actors = actor_cov)+
    remstats::send(variable = "size_medium", attr_actors = actor_cov)+
    remstats::send(variable = "size_large", attr_actors = actor_cov)+
    remstats::send(variable = "size_verylarge", attr_actors = actor_cov)+
    remstats::receive(variable = "IsPubHouse", attr_actors = actor_cov)+
    remstats::receive(variable = "BonPow",attr_actors =  actor_cov)+
    remstats::receive(variable = "size_medium", attr_actors = actor_cov)+
    remstats::receive(variable = "size_large", attr_actors = actor_cov)+
    remstats::receive(variable = "size_verylarge", attr_actors = actor_cov)+
    remstats::tie(attr_dyads = dyad_cov, variable = "distance",scaling="std")+
    remstats::tie(attr_dyads = dyad_cov, variable = "rivalry")+
    remstats::inertia(scaling = "prop")+ #FrPSndSnd
    remstats::rrankReceive() + #RRecSnd
    remstats::rrankSend() +#RSndSnd
    remstats::itp(scaling="std")+
    remstats::otp(scaling="std")+
    remstats::psABBA()+
    remstats::psABBY()+
    remstats::psABAY()

statistics_array <- remstats::tomstats(effects = statistics, reh = reh)

# Estimate the model
est <- remstimate::remstimate(reh = reh, stats = statistics_array, model = "tie", method = "MLE", ncores = 1)

params <- est$coefficients

# Simulate from this model
sim_statistics <- ~ remulate::baseline(param = params["baseline"])+
    remulate::send(param = params["send_BonPow"], variable = "BonPow", attributes = actor_cov)+
    remulate::send(param = params["send_size_medium"], variable = "size_medium", attributes = actor_cov)+
    remulate::send(param = params["send_size_large"], variable = "size_large", attributes = actor_cov)+
    remulate::send(param = params["send_size_verylarge"], variable = "size_verylarge", attributes = actor_cov)+
    remulate::receive(param = params["receive_IsPubHouse"], variable = "IsPubHouse", attributes = actor_cov)+
    remulate::receive(param = params["receive_BonPow"], variable = "BonPow",attributes =  actor_cov)+
    remulate::receive(param = params["receive_size_medium"], variable = "size_medium", attributes = actor_cov)+
    remulate::receive(param = params["receive_size_large"], variable = "size_large", attributes = actor_cov)+
    remulate::receive(param = params["receive_size_verylarge"], variable = "size_verylarge", attributes = actor_cov)+
    remulate::dyad(param = params["tie_distance"], attributes = dyad_cov, variable = "distance",scaling="std")+
    remulate::dyad(param = params["tie_rivalry"], attributes = dyad_cov, variable = "rivalry")+
    remulate::inertia(param = params["inertia"], scaling = "prop")+ #FrPSndSnd
    remulate::rrankReceive(param = params["rrankReceive"] ) + #RRecSnd
    remulate::rrankSend(param = params["rrankSend"] ) + #RSndSnd
    remulate::itp(param = params["itp"], scaling="std")+
    remulate::otp(param = params["otp"], scaling="std")+
    remulate::psABBA(param = params["psABBA"] )+
    remulate::psABBY(param = params["psABBY"] )+
    remulate::psABAY(param = params["psABAY"] )



# Simulate 100 relational event histories    
net_folder = "./net/"
runs = 100
for(r in 1:runs){
    dat <- remulate::remulateTie(effects = sim_statistics, actors = 1:33, time = 100000, events = nrow(edgelist))
    save(dat,file=paste0(net_folder,"dat_",r,".rdata"))
}

\end{lstlisting}
    
\end{appendices}


\end{document}